\documentclass[12pt,a4paper]{article}

\usepackage{amsmath}
\usepackage{mathrsfs}

\usepackage{harvard}
\usepackage{graphicx}

\usepackage{amssymb}
\begin{document}
\bibliographystyle{agsm}
\citationmode{abbr}
\citationstyle{agsm}
\title{\Large \textbf{Bayesian inference of nanoparticle-broadened
    x-ray line profiles} }
\author{N. Armstrong\footnote{E-mail:
    Nicholas.Armstrong@uts.edu.au} \footnote{Present address: Department of
    Applied Physics, University of Technology Sydney, PO Box 123, Broadway
    NSW 2007, AUSTRALIA.}, W. Kalceff$^\ddagger$, J. P. Cline  \& J. Bonevich}
\date{}
\maketitle
\pagestyle{myheadings}
\markright{\hspace{0mm} \footnotesize \emph Armstrong et al. (2001). ``Bayesian inference of nanoparticle-broadened
    x-ray line profiles'' }
\begin{center}
  National Institute of Standards and Technology,\\
Gaithersburg, Md 20899, USA\\
 $\ddagger$University of Technology Sydney,
\\PO Box 123, Broadway
    NSW 2007, AUSTRALIA
\end{center}

\abstract{A single and
self-contained method for determining the crystallite-size
distribution and shape from experimental x-ray line profile data is presented. We
have shown that the crystallite-size distribution can be
determined without assuming a functional form for the size
distribution, determining instead the size distribution with the
least assumptions by applying the Bayesian/MaxEnt method. The
Bayesian/MaxEnt method is tested using both simulated and experimental
CeO$_{2}$ data. The results demonstrate that the proposed method
can determine size distributions, while making the least number of assumptions.
The comparison of the Bayesian/MaxEnt results from experimental
CeO$_2$ with TEM results is favorable.\footnote{Invited talk and paper at \emph{Accuracy in Powder Diffraction III}, 22-25 April 2001, NIST, Gaithersburg, USA.}
}

\section{Introduction}\label{sec_intro}
The analysis of x-ray line profile  broadening can be considered
as solving a series of inverse problems. There are usually two
steps:--- removing the instrumental contribution (deconvolution),
and determining the broadening contribution in terms of
crystallite size and microstrain. Here we are concerned with
quantifying only the size broadening, in terms of the shape and size
distributions of the crystallites. We present a method that
removes the instrumental broadening and determines the particle
size distribution in a single step. The general theoretical
framework developed makes it possible to determine the crystallite
shape and average dimensions, and to fully quantify these results
by also assigning
 uncertainties to them.

There are two approaches that can be adopted. The first  assumes
functional forms for the size distribution and shape of the
crystallite, and applies a least squares fitting to determine the
parameters defining the size distribution
\cite{krill98,langford00}. For pragmatic reasons, this approach is
often used to ensure  numerical stability; however, it is based on
an explicit assumption for the crystallite size distribution and
does not take into account the uniqueness of the solution. 

The second approach takes into account the non-uniqueness of the
problem of determining the size distribution $P(\mathbf{D})$ from the experimental data,
by assigning a probability to the solutions and enabling an
average solution to be determined from the set of solutions;
moreover, it also allows any \emph{a priori} information and
assumptions to be included and tested. This approach is embodied
in the Bayesian and maximum entropy methods
\cite{gull89,skilling89a,bryan90a,sivia96}. Essentially, Bayesian
theory tells us how to express and manipulate probabilities. It
might be said, therefore, that Bayesian theory helps us to ask the
appropriate questions, while the maximum entropy method tells how
to assign values to quantities of interest.

\section{X-ray line profiles}\label{theory_broadening}

\subsection{Observed profile}
The observed line profile, $g(2\theta)$, can be expressed as
\begin{equation}
  g(2\theta)=\int\, k(2\theta
  -2\theta')\, f(2\theta')\,\mathrm{d}(2\theta') + b(2\theta) +
  n(2\theta) \label{equ_obs}
\end{equation}
where $k(2\theta)$ defines the instrument profile and considers
the imperfect optics of the diffractometer; $f(2\theta)$ is the
specimen profile, which (apart from strain effects which are not
covered here) characterizes the size broadening due to
microstructural properties of the specimen (i.e. crystallite
shape, distribution and dimensions); $b(2\theta)$ and $n(2\theta)$
are the background level and the noise distribution, respectively.
The observed profile, (\ref{equ_obs}), can also be expressed in
terms of reciprocal-space units, $s$, centered about
$s_{0}=\frac{2\sin\theta_{0}}{\lambda}$, as
\begin{equation}
  g(s)=g(2\theta)\,\frac{\mathrm{d}(2\theta)}{\mathrm{d}s} \label{equ_rspace}
\end{equation}
where
$\mathrm{d}(2\theta)=\frac{\lambda}{\cos\theta}\,\mathrm{d}s$.

The problem we face is determining the size distribution and shape
of the crystallites from (\ref{equ_obs}), given our knowledge of
the instrument kernel, $k(2\theta)$, and our understanding of the
counting statistics, $\sigma^{2}$. We also want to quantify the
specimen profile and size distribution by assigning error bars to
them. Before addressing these questions, we review line profile
broadening from nanocrystallites.

\subsection{Crystallite-size broadening}\label{sec_sizebroadening}
\begin{figure}[thb!]
  \hspace{20mm}\includegraphics[scale=0.7]{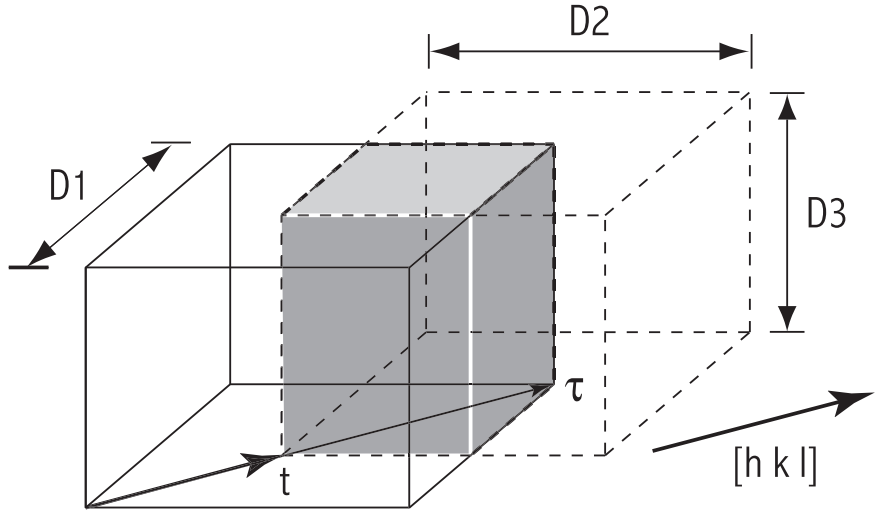}
  \caption{\footnotesize The crystallite (solid line) and its
    `ghost' (dashed line) shifted a
    distance $t$ in the direction of the scattering vector
    $[hkl]$. The crystallite and ghost have dimensions, $\mathbf{D}=\{D_{1},\,
    D_{2},\,D_{3}\}$. The shaded region represents the
    common volume between the crystallite and ghost. The maximum thickness of the
crystallite in the
    direction $[hkl]$ is $\tau$. The common-volume function has the
    boundary conditions, $V(0,\,\mathbf{D})=V_{0}$ and
    $V(\pm\tau,\,\mathbf{D})=0$. As $t\rightarrow \tau$,
    $V(t,\,\mathbf{D}) \rightarrow 0$. The
    graph of this function over $t$ represents the Fourier
    coefficients from which the area- and volume-weighted sizes can be
    determined.\label{fig_common}}
\end{figure}

The line profile, $I_{p}(s,\,\mathbf{D})$, from a specimen
consisting of crystallites of the same size and shape can be
expressed in terms of the common-volume function \cite{stokes42}
\begin{equation}
I_{p}(s,\,\mathbf{D})=2\int_{0}^{\tau}\,V(t,\,\mathbf{D})\,\cos
2\pi s t\,\mathrm{d}t \label{equ_intprofile}
\end{equation}
where $I_{p}(s,\,\mathbf{D})$ is the intensity profile given by
the dimensions of the crystallite,
$\mathbf{D}=\{D_{i};\,i=1,2,3\}$. The common-volume function of
the crystallite, $V(t,\,\mathbf{D})$, quantifies the volume
between the crystallite and its `ghost', shifted a distance $t$
parallel to the diffraction vector.  The dimension $\tau$
represents the maximum length of the crystallite in the direction
of the diffraction vector, and can be expressed in terms of the
dimensions of the crystallite, $\mathbf{D}$, such that
$\tau\equiv\tau(\mathbf{D})$. The boundary conditions for the
common-volume function are $V(0,\,\mathbf{D})=V_{0}$, where
$V_{0}$ is the volume of the crystallite, and
$V(\pm\tau,\,\mathbf{D})=0$. Fig.~\ref{fig_common} shows a
schematic of a crystallite and its ghost shifted a distance $t$ in
the direction $[hkl]$; the shaded region represents the common
volume between the crystallite and its ghost. $V(t,\,\mathbf{D})$
is symmetrical about the origin over the range
$t\in[-\tau,\,\tau]$. This implies that $V(t,\,\mathbf{D})$ is an
even function over this range. A simple example is a set of spherical crystallites with diameter $D$, for which the common-volume is given by \cite{stokes42}
\begin{subequations}
\begin{equation}
V(t,\,D)=\frac{\pi}{12} (t+2\,D)(t-D)^{2} \label{equ_commonsphere}
\end{equation}
and using (\ref{equ_intprofile}) the corresponding line-profile is \cite{stokes42,langford00}
\begin{equation}
I_{p}(s,\,D)=\frac{1}{16 \pi^{3} s^{4}} + \frac{D^{2}}{8 \pi
      s^{2}} - \frac{\cos(2\pi s D)}{16 \pi^{3}
      s^{4}} - \frac{D\, \sin(2\pi s
      D)}{8 \pi^{2} s^{3}} \label{equ_sphereprofile}
\end{equation}\label{equ_sphereexample}
\end{subequations}
where $\tau(D)=D$ for spherical crystallites and in the limit of $s\rightarrow 0$ (\ref{equ_sphereprofile}) reduces to $I_{p}(0,\,D)=\pi D^{4}/8$.

Essentially, (\ref{equ_intprofile}) is the Fourier transform of
the $V(t,\,\mathbf{D})$, and noting $V(t,\,\mathbf{D})$ is an even
function, the odd  (sine) terms in the Fourier transform vanish.
This also implies that the size-broadened  profiles will always be
symmetrical about the Bragg angle, $2\theta_{0}$. From
(\ref{equ_intprofile}) and Fig.~\ref{fig_common}, it is clear that
information concerning the dimensions and shape  of the
crystallite is given in $V(t,\,\mathbf{D})$.

\subsection{Particle-size distribution, $P(\mathbf{D})$}
A powder specimen would not normally consist of crystallites all
having the same size, but it can be assumed that the
crystallites can have the same shape, based on kinetics arguments.
The effect of the particle-size distribution on the common volume
is to `blur' the broadening effects of a single crystallite.

The size-broadened line profile from a distribution of
crystallites, $P(\mathbf{D})\, \mathcal{D}\mathbf{D}$, with
dimensions in the range  $\mathbf{D}$ to
$\mathbf{D}+\mathcal{D}\mathbf{D}$ can be expressed as
\begin{equation}
  f(s) =2 \int_{0}^{\infty}\, \tilde{V}(t)\,\cos 2\pi s
  t\,\mathrm{d}t, \,\,\,\, \, \, \forall \, s\in [-\infty,\,+\infty]\label{equ_sprofile}
\end{equation}
where $\tilde{V}(t)$ is the \emph{modified} common-volume function
due to the influence of the particle-size distribution,
\begin{equation}
   \tilde{V}(t)=\int_{t}^{\infty}\,V(t, \,
   \mathbf{D})\,P(\mathbf{D})\, \mathcal{D}\mathbf{D}.
   \label{equ_modcomvol}
\end{equation}
In (\ref{equ_modcomvol}) a generalized measure,
$\mathcal{D}\mathbf{D}$, has been used which is dependent on the
crystallite shape and coordinate system. The area-weighted size,
$\langle t \rangle_{a}$, volume-weighted size,$\langle t
\rangle_{v}$, and column-length distribution (or area-weighted
size distribution), $p_{a}(t)$, can be determined from
(\ref{equ_modcomvol}) \cite{wilson68a,wilson71a}. It can be seen
from (\ref{equ_modcomvol}) how the shape and distribution of the
crystallites influence the area- and volume-weighted quantities.
Substituting (\ref{equ_modcomvol}) into (\ref{equ_sprofile}), we
have
\begin{subequations}
  \begin{eqnarray}
  f(s) &=& 2\int_{0}^{\infty}\left[\int_{t}^{\infty}\,V(t,\,\mathbf{D})\,
    P(\mathbf{D})\,\mathcal{D}\mathbf{D}\right]\, \cos 2\pi s t \,\mathrm{d}t
    \label{equ_sprofile2}\\
    &=& \int_{0}^{\infty}\,\left[2\int_{0}^{\tau}\,V(t,\,\mathbf{D}) \cos
  2\pi s t \,\mathrm{d}t \right]\,
  P(\mathbf{D})\,\mathcal{D}\mathbf{D} \label{equ_sprofile3}
  \end{eqnarray}
\label{equ_sizeprofiles}
\end{subequations}
where in going from (\ref{equ_sprofile2}) to (\ref{equ_sprofile3})
the order of integration has been changed and $t$ is integrated
out. In addition we note that $V(t,\,\mathbf{D}) \geq 0$ for $t\in [0,\,\tau]$ and $V(t,\,\mathbf{D})= 0$ for $t>\tau$. Inside the brackets of (\ref{equ_sprofile3}), we have
$I_{p}(s,\,\mathbf{D})$ from (\ref{equ_intprofile}). Hence,
(\ref{equ_sprofile3}) can be written as
\begin{equation}
  f(s)=\int_{0}^{\infty}\,
  I_{p}(s,\,\mathbf{D})\,P(\mathbf{D})\,\mathcal{D}\mathbf{D},\,\,\,\, \, \, \forall \, s\in [-\infty,\,+\infty] 
  \label{equ_sprofile4}
\end{equation}
where we define the profile kernel, $I_{p}(s,\,\mathbf{D})$, as
the size-broadened line profile given by a single crystallite with
dimensions $\mathbf{D}$. In (\ref{equ_sprofile4}), we notice that
the effect of $P(\mathbf{D})$ is to weight the superposition of
size profiles over the range of $\mathbf{D}$ to
$\mathbf{D}+\mathcal{D}\mathbf{D}$.
\subsection{Determining $P(\mathbf{D})$ from $g(s)$}\label{sec_g}
In analysing the size distribution, we want to ensure that the
statistics of the observed profile can be carried directly into
quantifying the size distribution. Equation (\ref{equ_sprofile4})
expresses the specimen profile, $f(s)$, in terms of the
particle-size distribution and the shape of the nanocrystallites,
while (\ref{equ_obs}), after transformation into $s$-space,
expresses the observed profile in terms of $f(s)$. Combining these
two equations, the experimental data, $g(s)$, can be expressed in
terms of the particle-size distribution, $P(\mathbf{D})$ as
\begin{subequations}
\begin{eqnarray}
  g(s) &=& \int_{0}^{+\infty}
  \int_{-\infty}^{+\infty}\,k(s-s')\,I_{p}(s',\,\mathbf{D})\,P(\mathbf{D})\,\mathrm{d}s'
  \mathcal{D}\mathbf{D} + b(s) + n(s)
  \label{equ_allinone}\\
       &=& \int_{0}^{+\infty}\,
       K(s,\,\mathbf{D})\,P(\mathbf{D})\,\mathcal{D}\mathbf{D}  + b(s) +
       n(s)\label{equ_allinone2}
\end{eqnarray} \label{equ_single}
\end{subequations}
where the scattering kernel, $K(s,\,\mathbf{D})$, `rolls up' the
instrumental effects and the profile kernel, given by
\begin{equation}
K(s,\,\mathbf{D})=\int_{-\infty}^{+\infty}\,
k(s-s')\,I_{p}(s',\,\mathbf{D})\,\mathrm{d}s'.
  \label{equ_kernel}
\end{equation}
In (\ref{equ_kernel}), the dummy variable $s'$ is being integrated
out. The results given by (\ref{equ_allinone2}) and
(\ref{equ_kernel}) enable the particle-size distribution to be
extracted directly from the experimental data. This ensures that
the statistics of the experimental data are transferred to
quantifying the uncertainty in the solution. This approach also
addresses a difficulty of the \emph{two-fold approach} discussed
by \citeasnoun{armstrong99b}.

\section{Bayesian \& maximum entropy methods}\label{sec_bayes}

\subsection{The uniqueness of $P(\mathbf{D})$}\label{sec_uniqueness}
In (\ref{equ_single}) we have a single expression for the observed
profile in terms of the crystallite size distribution and shape,
background level, and statistics of the experiment; information
concerning the crystallite properties has been incorporated.

In seeking to determine $P(\mathbf{D})$ from $g(s)$, the issue of
uniqueness for $P(\mathbf{D})$ becomes important, for two reasons:
firstly, because of the `conditioning' of the kernels,
particularly $K(s,\,\mathbf{D})$; and secondly, due to the
presence of statistical noise, $\sigma$.

Generally, $K(s,\,\mathbf{D})$ will be \emph{ill-conditioned}.
This can be demonstrated in a numerical calculation by expressing
$K(s,\,\mathbf{D})$ as a matrix, $\mathbf{K}$; we can show $\det
\mathbf{K}^{T} \mathbf{K} \sim 0$. This implies that the column
vectors of  $\mathbf{K}$ are (nearly all) linearly dependent,
which has dire consequences, as any attempt to determine
$P(\mathbf{D})$ (given $g(s)$, $K(s,\,\mathbf{D})$, $\sigma$ and
$b(s)$), produces a set of solutions $\{P(D)\}$ rather than a
unique solution.  The presence of statistical noise in the data
simply worsens the situation, in that the ill-conditioning of
$K(s,\,\mathbf{D})$ amplifies the noise and the solution is
swamped by spurious and unphysical oscillations
\cite{armstrong98a}. Faced with this situation, the following
question arises:

\emph{How do we develop a method to extract a unique
$P(\mathbf{D})$ from $g(s)$, given our knowledge of
$K(s,\,\mathbf{D})$, $b(s)$ and $\sigma^{2}$?}

\subsection{Some observations}\label{sec_observations}
Before proceeding with developing a `method' to determine the
crystallite size and shape from the observed data, $g(s)$, some
observations concerning these distributions need to be made.

The integral equations given by (\ref{equ_obs}) and
(\ref{equ_single}) refer to a set of continuous functions.
However, the recording of the observed and instrument profiles is
made in discrete time intervals. To convey this, we express the
observed profile, specimen profile and size distribution as 
vectors, such that
$\mathbf{g}=\{g_{i};\,i=1,\,2,\,3,\,\ldots,\,M\}$,
$\mathbf{f}=\{f_{j'};\,j'=1,\,2,\,3,\,\ldots,\,N'\}$ and
$\mathbf{P}=\{P_{j};\,j=1,\,2,\,3,\,\ldots,\,N\}$. The
scattering kernel $K(s,\,\mathbf{D})$ can be expressed as matrix,
$\mathbf{K}=\{K_{ij}; \, \forall \ i \,\& \,j\}$ by taking the
product of the instrument kernel and the line profile kernel. The
instrument kernel can be evaluated in $2\theta$-space, such that
$\mathbf{R}=\{k(2\theta_{i}-2\theta_{j'}');\,i=1,\,2,\,3,\,\ldots,\,M
\, \& \, j'=1,\,2,\,3,\,\ldots,\,N' \}$, and using
$\mathrm{d}(2\theta)=\frac{\lambda}{\cos\theta}\,\mathrm{d}s$ can
be mapped into $s$-space. Similarly, the profile kernel can be
evaluated over $s$ and $\mathbf{D}$, such that
$\mathbf{I}_{p}=\{I_{p\,j'j};\,j'=1,\,2,\,3,\,\ldots,\,N' \, \& \,
j=1,\,2,\,3,\,\ldots,\,N\}$. The matrix product gives
$\mathbf{K}=\mathbf{R}\,\mathbf{I}_{p}$ and is an $[M \times
N]$ matrix, such that  $N < N' \leq M$.

There are \emph{two} fundamental properties which $g(2\theta)$,
$f(2\theta) $, $P(\mathbf{D})$ and $V(t,\,\mathbf{D})$ all share.
The first is that these distributions are \emph{positive}
definite; that is, the observed profile $g(2\theta)$ and specimen
function $f(2\theta)$ represent intensities which are positive values.
The second property is that these distributions are
\emph{additive}; that is, the sum of the distributions over a
region represents a physically meaningful quantity \cite{sivia96}.
For example, the integrated intensity of $g(s)$ can be related
back to the structure factor of the lattice, while the integrals
$\int f(s)\,\mathrm{d}s$ and $\int V(t,\,\mathbf{D})\,\mathrm{d}t$
are inversely proportional to the integral breadth and quantify
the specimen broadening in terms of size and strain contributions.
The integral for $P(\mathbf{D})$ is a special case, in that it 
must be unity. This ensures that we can attribute a
probability for a particular $\mathbf{D}$ and detemine its
moments.

These two observations are important in formulating a `method'
that can determine both the specimen profile from the observed
x-ray diffraction profile and an underlying distribution such as
the size distribution, $P(\mathbf{D})$, while dealing with the
issue of uniqueness. That is, we expect our method to extract this
information from the observed data and produce results which
preserve the positivity and additivity of the profile or
distribution. It should also be possible to incorporate the
properties of positivity and additivity without making additional
assumptions about, say,  the functional/analytical form of the
specimen profile or size distribution. These conditions ensure
that the specimen profile or size distribution determined from the
observed profile can be interpreted in general terms.

In order to assign values to these distributions and preserve
their additivity and positivity, a suitable function must be
selected. Based on these observations and various arguments, the
entropy function and maximum entropy principle are found to be the
\emph{only} consistent approach to inferring discrete
probabilities \citeaffixed{shore80,johnson83,tikochinsky84,gull84,skilling89b,skilling90,goambo97}{see}.

\subsection{Bayes' theorem for $P(\mathbf{D})$}
In analyzing size-broadened profiles, the central aim is to
quantify the shape and size distribution of the crystallites,
given the experimental data. Bayesian theory is well suited for
testing a \emph{hypothesis} in the presence of experimental data.
This is achieved by  quantifying the \emph{a posteriori}
probability distribution for $\mathbf{P}$, conditional on the
experimental data and statistical noise. The formulation of Bayes'
theorem is general and can also be applied to determining
$\mathbf{f}$.

Using Bayes' theorem, the  \emph{a posteriori} probability for
$\mathbf P$ is given by
\begin{equation}
\Pr(\mathbf{P}|\,\mathbf{g},\,\mathbf{m},\mathbf{K},\,\sigma,\alpha,\,\mathcal{I})
=\frac{\Pr(\mathbf{P}|\,\mathbf{m},\alpha,\,\mathcal{I})\,\Pr(\mathbf{g}|\,\mathbf{P},\,\mathbf{K},\,\sigma,\,\mathcal{I})}{\Pr(\mathbf{g}|\,\mathbf{m},\,\mathbf{K},\,\sigma,\,\mathcal{I})}
  \label{equ_btheorem}
\end{equation}
This is conditional on everything after `$|$', viz.  the observed
profile $\mathbf{g}$, an \emph{a priori} model $\mathbf{m}$, the
scattering kernel $\mathbf{K}$, statistical noise $\sigma$, a constant
$\alpha$, and any additional background information concerning the
experiment, $\mathcal I$.

On the right-hand side of (\ref{equ_btheorem}) there are several
terms that require further discussion.

The \emph{likelihood} probability distribution
$\Pr(\mathbf{g}|\,\mathbf{P},\,\mathbf{K},\,\sigma,\,\mathcal{I})$
defines the probability of measuring $\mathbf{g}$, given a size
distribution $\mathbf{P}$, profile kernel $\mathbf{K}$, and
statistical noise $\sigma$. That is, we include our hypothesis
$\mathbf{P}$, and determine how probable it is to measure
$\mathbf{g}$, given this hypothesis, $\mathbf{K}$ and $\sigma$.
The likelihood function is approximated as a Gaussian distribution
for large counts ($>>10$) by applying the \emph{central limit
theorem},
\begin{subequations}
\begin{equation}
  \Pr(\mathbf{g}|\,\mathbf{P},\,\mathbf{K},\,\sigma,\,\mathcal{I})=
 \frac{1}{Z_{L}(\sigma)}\,\exp[-\frac{1}{2}\,L(\mathbf{P},\,\mathbf{g},\,\mathbf{K},\,
 \sigma)]
  \label{equ_likelihood}
\end{equation}
where
\begin{equation}
L=\sum_{i=1}^{M}\,\frac{\left(g_{i}-\sum_{j=1}^{N}\,K_{ij}\,P_{j}\right)^{2}}{\sigma_{i}^
{2}} \label{equ_chisq}
\end{equation}
and
\begin{eqnarray}
  Z_{L}(\sigma) &=& \prod_{i=1}^{M}\,\sqrt{2 \pi\,\sigma_{i}^{2}}
  \label{equ_liknorm} \\
                &=& \det \{\sqrt{2 \pi\,\sigma^{2}}\},\label{equ_liknorm2}
\end{eqnarray}
\end{subequations}
such that $\{\sqrt{2 \pi\,\sigma^{2}}\}$ is an $[M \times M]$
diagonal matrix.

The variance is defined in terms
of the observed counts and estimated background-level as
$\sigma_{i}^{2} = g_{i} + b_{i}^{est}$. In
(\ref{equ_likelihood}), the kernel, $\mathbf{K}$ has been included
as it contains information about the shape of the crystallites and
will influence the solution. We notice from (\ref{equ_chisq}) that
the matrix form of (\ref{equ_single}) has been incorporated.

The term $\Pr(\mathbf{P}|\,\mathbf{m},\alpha,\,\mathcal{I})$
defines how probable is our hypothesis $\mathbf{P}$, given it is a
positive and additive distribution and conditional on an \emph{a
priori} model, $\mathbf m$. The \emph{a priori} probability
distribution can be expressed as
\begin{subequations}
\begin{eqnarray}
  \Pr(\mathbf{P}|\,\mathbf{m},\alpha,\,I) &=&
  \frac{1}{Z_{S}(\alpha)}\,\exp\left[\alpha\,S(\mathbf{P},\,\mathbf{m})\right]
  \label{equ_apriori}.
\end{eqnarray}
The entropy function is given as \cite{skilling89a},
\begin{equation}
  S(\mathbf{P},\,\mathbf{m}) = \sum_{j=1}^{N} \,P_{j} - m_{j}-P_{j}
  \ln\left(P_{j}/m_{j} \right).\label{equ_entropy}
\end{equation}
where the normalization term, $Z_{S}(\alpha)$ is given as
\begin{eqnarray}
  Z_{S}(\alpha)&=&\int\mathcal{D}\mathbf{P}\,
  \exp\left[\alpha\,S(\mathbf{P},\,\mathbf{m})\right] \label{equ_zalpha1}\\
               &=& \left(\frac{2 \pi}{\alpha}\right)^{\frac{N}{2}}
  \label{equ_zalpha2}\\
               &=& \frac{(2 \pi)^\frac{N}{2}}{
               \sqrt{\det \alpha \mathbf{I}}} \label{equ_zalpha3}
\end{eqnarray}
and the integration in (\ref{equ_zalpha1}) involves the measure
$\mathcal{D}\mathbf{P}=
\prod_{j=1}^{N}\,P_{j}^{-\frac{1}{2}}\,\mathrm{d}P_{j}$.

\end{subequations}
The log term in (\ref{equ_entropy}) ensures that positive and
additive distributions are obtained and that $\mathbf P$ will have
these fundamental characteristics. The \emph{a priori model},
$\mathbf m$, defines our ignorance/knowledge about $\mathbf P$.
That is, if we are unsure of the shape of $\mathbf P$, it is best
to admit our ignorance by assigning a uniform distribution over a
specified range. The \emph{a priori} model may also include data
gathered from other sources, such as electron microscopy (e.g.
TEM, SEM and SPTM) techniques. It may also include theoretical or
analytical models. For example, recently in the literature
\citeaffixed{krill98,langford00,ungar01}{see} there has been a
widespread use of the log-normal distribution for $\mathbf P$.
However, in the Bayesian formulation we \emph{do not explicitly}
define $\mathbf P$ as a log-normal distribution, but set the
\emph{a priori} model as a log-normal distribution and test it in
the presence of the observed data.

$S(\mathbf{P},\,\mathbf{m})$ is essentially a measure for $\mathbf
P$ relative to $\mathbf m$. Suppose the model $\mathbf{m}$ was
found to be a log-normal distribution and its parameters
determined using least squares analysis. If the resulting $\mathbf P$ lies
`close' to $\mathbf m$, the change in $S$ will be small and there
will be little difference between $\mathbf{P}$ and $\mathbf{m}$; also, this would imply that the underlying crystallite-size
distribution in the specimen is a log-normal distribution with
values similar to those determined for $\mathbf m$, since this
assumption has been tested in the presences of the experimental
data. On the other hand, if $\mathbf P$ lies `some distance' from
$\mathbf m$, the change in $S$ will be large; this would result in
a considerable difference between $\mathbf{m}$ and $\mathbf{P}$ and would imply that the underlying size distribution
is not a log-normal distribution with the values estimated for
$\mathbf m$.

The denominator term in (\ref{equ_btheorem}) has an important
application in selecting between various kernels, $\mathbf{K}$,
for different crystallite shapes. It is called the \emph{evidence}
\cite{gull89},
\begin{equation}
  \Pr(\mathbf{g}|\,\mathbf{m},\,\mathbf{K},\,\sigma,\,\mathcal{I}) =
  \int\mathcal{D}\mathbf{P}\,\int\mathrm{d}\alpha\,
  \Pr(\mathbf{P},\,\mathbf{g},\,\alpha|\,\mathbf{m},\,
  \mathbf{K},\,\sigma,\,\mathcal{I}).
  \label{equ_evid}
\end{equation}

Including all the necessary terms, the \emph{a posteriori}
probability distribution for $\mathbf{P}$ can be expressed as
\begin{equation}
  \Pr(\mathbf{P}|\,\mathbf{g},\,\mathbf{m},\mathbf{K},\,
  \sigma,\alpha,\,\mathcal{I})
  = \frac{1}{Z_{S}(\alpha)\,Z_{L}(\sigma)}\,
  \frac{e^{Q} }{\Pr(\mathbf{g}|\,\mathbf{m},\,\mathbf{K},\,
  \sigma,\,\mathcal{I})}
  \label{equ_btheorem2}
\end{equation}
where $Q =\alpha\,S-\frac{1}{2}\,L$. For convenience, $Q\equiv
Q(\mathbf{P},\,\alpha)$, since $\mathbf{P}$ and $\alpha$ are the
only two unknown terms. The $\alpha$ term in $Q(\mathbf{P},\,
\alpha)$ can be interpreted as an undetermined Lagrangian
multiplier.

Determining the most probable size distribution,
$\hat{\mathbf{P}}$, depends on maximizing (\ref{equ_btheorem2}),
which in turn requires determining the global minimum for
$Q(\mathbf{P})$. There are several algorithms for determining
$\hat{\mathbf{P}}$ from $Q(\mathbf{P})$, given its nonlinear
characteristics \citeaffixed{skilling84,bryan90a}{see}.

The approach we follow in determining the crystallite-size
distribution is similar to that outlined by \citeasnoun{bryan90a}
and \citeasnoun{jarrell96}. We start with a large $\alpha$ value
and step towards $\alpha \approx 0$. For a given $\alpha$, we
determine $\mathbf{P}$ such that $\nabla Q =0$. After stepping
through a range of $\alpha$ values, a set of solutions,
$\{\mathbf{P}(\alpha)\}$, is formed parameterized by $\alpha$.
The average distribution, $\langle \mathbf{P} \rangle$, can be
determined from the set of solutions $\{\mathbf{P}(\alpha)\}$,
\begin{equation}
  \langle \mathbf{P} \rangle =
\int_{\alpha_{min}}^{\alpha_{max}}\mathrm{d}\alpha\,\mathbf{P}(\alpha)\,\Pr(\alpha|\,\mathbf{g},\,
\mathbf{m},\,\mathbf{K},\,\sigma,\,\mathcal{I}).
  \label{equ_avesize}
\end{equation}
where
$\Pr(\alpha|\,\mathbf{g},\,\mathbf{m},\,\mathbf{K},\,\sigma,\,\mathcal{I})$
is normalized to unity for $\alpha \in
[\alpha_{min},\,\alpha_{max}]$. In the application of the
Bayesian/MaxEnt method, the selected range was defined by $\alpha \in
[10^{-2},\,10^{5}]$. The average particle size
distribution can be used to determine the average specimen
profile, $\langle \mathbf{f} \rangle$,
\begin{eqnarray}
  \langle \mathbf{f} \rangle &=&
  \int_{\alpha_{min}}^{\alpha_{max}}\mathrm{d}\alpha\,\mathbf{I}_{p}
  \mathbf{P}(\alpha)\,\Pr(\alpha|\,\mathbf{g},\,\mathbf{m},\,
  \mathbf{K},\,\sigma,\,
\mathcal{I})\nonumber\\
  &=& \mathbf{I}_{p}\, \langle \mathbf{P}
  \rangle. \nonumber \label{equ_aveprofile}
\end{eqnarray}

\subsection{Determining $\Pr(\alpha|\mathbf{g},\,\mathbf{m},\,
\mathbf{K},\,\sigma,\,\cal{I})$} 
The $\alpha$ parameter in
(\ref{equ_btheorem2}) is important in coupling the  entropy
function $S(\mathbf{P},\,\mathbf{m})$ with the likelihood
function $L(\mathbf{P})$. It is also a `nuisance parameter' and
its influence can be integrated out. In evaluating
(\ref{equ_avesize}), it is necessary to determine
$\Pr(\alpha|\,\mathbf{g},\,\mathbf{m},\,\mathbf{K},\,
\sigma,\,\mathcal{I})$; we do this by integrating out the
influence of $\mathbf P$,
\begin{subequations}
 \begin{eqnarray}
    \Pr(\alpha|\,\mathbf{g},\,\mathbf{m},\,\mathbf{K},\,\sigma,\,\mathcal{I})
&=&
\int\mathcal{D}\mathbf{P}\,\Pr(\mathbf{P},\,\alpha|\,\mathbf{g},\,\mathbf{m},\,\mathbf{K},\,\sigma,\,\mathcal{I})
\nonumber \\
&=& \int\mathcal{D}\mathbf{P}\,\frac{\Pr(\alpha|\,\mathcal{I})\,\Pr(\mathbf{P}|\,\mathbf{m},\alpha,\,\mathcal{I})\,\Pr(\mathbf{g}|\,\mathbf{P},\,\mathbf{K},\,\sigma,\,\mathcal{I})}{\Pr(\mathbf{g}|\,\mathbf{m},\,\mathbf{K},\,\sigma,\,\mathcal{I})}\nonumber \\
&=&
 \frac{\Pr(\alpha|\,\mathcal{I})}{\Pr(\mathbf{g}|\,\mathbf{m},\,\mathbf{K},\,\sigma,\,
 \mathcal{I})}\,\frac{1}{Z_{S}(\alpha) \, Z_{L}(\sigma)}\nonumber\\
&\times&\int\mathcal{D}\mathbf{P}\,e^{Q(\mathbf{P},\,\alpha)},
   \label{equ_alphaint}
 \end{eqnarray}
and expanding $Q(\mathbf{P},\,\alpha)\approx
   Q(\hat{\mathbf{P}},\,\alpha)+\frac{1}{2}(\mathbf{P}-\hat{\mathbf{P}})
   ^{T}\nabla\nabla
   Q (\mathbf{P}-\hat{\mathbf{P}})$ about $\hat{\mathbf{P}}$ for a given
   $\alpha$. We note $\nabla Q=0$ for
   $\mathbf{P}=\hat{\mathbf{P}}$ for a given $\alpha$.
   On integrating, we have
\begin{eqnarray}
\Pr(\alpha|\,\mathbf{g},\,\mathbf{m},\,\mathbf{K},\,\sigma,\,
\mathcal{I}) &\approx&
\frac{\Pr(\alpha|\,\mathcal{I})}{\Pr(\mathbf{g}|\,\mathbf{m},\,
\mathbf{K},\,\sigma,\,
\mathcal{I})}\,\frac{1}{Z_{S}(\alpha)\, Z_{L}(\sigma)} \nonumber\\
&\times& \frac{
              (2\pi)^{Nq
              /2}\,e^{Q(\hat{\mathbf{P}},\,\alpha)}
               }{
                 \sqrt{\det \nabla \nabla Q(\alpha)}
                 }
                  \label{equ_alphaint2}\\
&=&\frac{
        \Pr(\alpha|\,\mathcal{I})
        }{
          \Pr(\mathbf{g}|\,\mathbf{m},\,\mathbf{K},\,\sigma,\,\mathcal{I})
          }\,\frac{1}{Z_{L}(\sigma)}
\nonumber\\
&\times& \sqrt{\frac{\det \alpha \mathbf{I}}{\det (\alpha
\mathbf{I} + \hat{\Lambda})}}\,e^{Q(\hat{\mathbf{P}},\,\alpha)}
\end{eqnarray}\label{equ_subequ1}
\end{subequations}
where $\nabla \nabla Q(\hat{\mathbf{P}},\,\alpha) \equiv \nabla
\nabla Q(\alpha)$ and $\hat{\Lambda}$ are the eigenvalues of
$(-\nabla\nabla S)^{-\frac{1}{2}}\,\nabla\nabla L \,(-\nabla\nabla
S)^{-\frac{1}{2}} = \{\hat{\mathbf{P}}^{\frac{1}{2}}\}
\,\mathbf{K}^{T}\{\sigma^{-2}\}\mathbf{K}\,\{\hat{\mathbf{P}}^{\frac{1}{2}}\}$.
The quantities in parentheses represent diagonal matrices. In
(\ref{equ_alphaint}), we have introduced the \emph{a priori}
distribution for $\alpha$, $\Pr(\alpha|\,\mathcal{I})$. Generally,
we set $\Pr(\alpha|\,\mathcal{I})$ as a uniform model over a range
$[\alpha_{min},\,\alpha_{max}]$. Using (\ref{equ_subequ1}) we can
evaluate (\ref{equ_avesize}). In practice, we determine $\ln
\Pr(\alpha|\,\mathbf{g},\,\mathbf{m},\,\mathbf{K},\,\sigma,\,\mathcal{I})$
and $\hat{\Lambda}$ for each $\hat{\mathbf{P}}$ and $\alpha$ in
the range of $[\alpha_{min},\,\alpha_{max}]$.

\subsection{Resolving overlapped profiles}
The formalism presented here enables single and overlapped profiles,
and even whole
patterns to be analyzed, provided that  crystallite-size effects
are the major broadening component. Line profiles are generally
overlapped due to low unit cell symmetry. However, specimen
broadening, such as size broadening from crystallites, can also cause
profiles to be overlapped. In this case, the underlying invariant
quantity is the crystallite-size distribution, $\mathbf{P}$. The
above integral equations for overlapped peaks can be
expressed in terms of $\mathbf P$. The general form of
(\ref{equ_single}) does not change; the term that does change is
the kernel, $\mathbf{K}(s,\,\mathbf{D})$,
\begin{equation}
  K(s,\,\mathbf{D})=
  \int_{-\infty}^{+\infty}\,\sum_{q}\,k(s-s';\,s'_{0q})\, I_{p}(s',
  \,\mathbf{D}) \,\mathrm{d}s' \label{equ_newkernel2}
\end{equation}
where $s'_{0q}=2\sin\theta_{0q}/\lambda$ and $\theta_{0q}$ is the
Bragg angle at the $q$th peak in the pattern. The
$k(s-s';\,s'_{0q})$ term expresses the instrument kernel at each
peak position, $\theta_{0q}$. The $I_{p}(s',\,\mathbf{D})$ term
is invariant over the range of $s$. In terms of the Bayesian
analysis presented above, nothing else changes.

\subsection{Error analysis}\label{sec_errorbars}
Determining the errors in $\mathbf{P}$ over regions of importance
is a final test for the quality of $\mathbf{P}$. The error bars
for  $\mathbf{P}$ are dependent on the choice of the \emph{a
priori} model and the quality of the observed data, $\sigma$.

It is only possible to assign error bars over a defined region,
because the errors between points are strongly correlated
\cite{skilling89a,sivia96}. The region of interest may consist of
features in the specimen profile or size distribution which may not
be physical, such as ripples in the tails of the
distribution or a second peak suggesting a bimodal distribution.
Over the defined region, we are interested in the \emph{average
integrated flux} \cite{skilling89a},
\begin{eqnarray}
  \rho &=& \sum_{j=1}^{N} P_{j} w_{j}/ \sum_{j=1}^{N} w_{j} w_{j} \label{equ_flux1}\\
       &=& \mathbf{P}^{T}\,\mathbf{w}/\mathbf{w}^{T}\mathbf{w} \label{equ_flux2}
\end{eqnarray}
where $\mathbf{w}$ is a `window function' defined as,
\begin{equation}
  w_{j}=\left\{\begin{array}{ll}
                        1 & \mbox{$r \leq j \leq r'$}, \\
                        0, & \mbox{otherwise}
                  \end{array}
            \right. \label{equ_win}
\end{equation}
and the region of interest is defined by $rr'$. Expanding
$\Pr(\mathbf{P}|\,\mathbf{g},\,\mathbf{m},\,\mathbf{K},\,\sigma,\,\alpha,\,\mathcal{I})$
about $\hat{\mathbf{P}}$, we have
$\Pr(\mathbf{P}|\,\mathbf{g},\,\mathbf{m},\,\mathbf{K},\,\sigma,\,\alpha,\,\mathcal{I})
\propto
e^{\frac{1}{2}(\mathbf{P}-\hat{\mathbf{P}})^{T}\nabla\nabla
   Q (\mathbf{P}-\hat{\mathbf{P}})}$. This is a Gaussian centered about
   $\hat{\mathbf{P}}$. By inspection, the covariance matrix for $\mathbf{P}$
   is given by $-(\nabla \nabla Q)^{-1}$, where the elements in $-(\nabla
   \nabla Q)^{-1}$ are strongly correlated with neighboring
   elements. Following the suggestion of \citeasnoun{skilling89a}
the variance for $\mathbf{P}$ is
\begin{equation}
  \sigma_{\mathbf{P}}^{2}= \mathbf{w}^{T} \left[-(\nabla \nabla
    Q)^{-1}\right]\mathbf{w}/ \mathbf{w}^{T}  \mathbf{w}\label{equ_var}.
\end{equation}
Hence, we can assign error bars over a region of interest to the
integrated flux of $\mathbf{P}$.

\subsection{Fuzzy pixel approach for determining
  $\mathbf{f}$ }\label{sec_fuzzy}
It is often important to assess the specimen broadening by
determining $\mathbf{f}$, without making any assumptions concerning its
functional form. This can be performed by deconvolving (\ref{equ_obs}).
However, in determining $\mathbf{f}$ `ringing effects' can appear
in the solution. The ringing is often due to noise effects which
are amplified and appear as unphysical oscillations in the
solution \citeaffixed{armstrong99b}{for example see Fig. 6 in}.
The above theory assumes that smoothing is applied
globally. However, the ringing effects are local artifacts. In
order to introduce `local' smoothing, we must address how to
decompose $\mathbf f$.  Explicit in the composition of
$\mathbf{f}$ is that it is expressed as a superposition of delta
functions,
\begin{equation}
  f(2\theta)=\sum_{l=1}^{N}\,\delta(2\theta-2\theta_{l}) \,a_{l} \label{equ_delta}
\end{equation}
where $\mathbf{a}=\{a_{1},\,a_{2},\,\ldots,\,a_{N}\}$ is the set
of coefficients that define the amplitude of $f$ at the $l$th
position. However, (\ref{equ_delta}) assumes a global smoothness,
while the ringing effects are local effects.

Following the suggestion of \citename{sivia99}
\citeyear{sivia99,sivia96}, we blur $\delta(2\theta)$ by including
the spatial correlation length or width. To do this, we choose a
basis function which includes a spatial correlation length as its
width and reduces to $\delta(2\theta)$ in the limit of the width
going to zero. That is, we make the pixel at the $l$th position of
$\mathbf f$ \emph{fuzzy}. A simple choice  is to express $\mathbf
f$ in terms of a sum of Gaussian function,
\begin{equation}
f(2\theta)=\sum_{l=1}^{N}\,\exp\left[-\frac{(2\theta-2\theta_{l})^{2}}{2\,\omega^{2}}
\right]\,a_{l}
  \label{equ_fuzzy}
\end{equation}
where $\omega$ is the width of the spatial correlation or fuzzy
pixel. In the limit of $\omega \rightarrow 0$, (\ref{equ_fuzzy})
reduces to (\ref{equ_delta}).

In matrix notation (\ref{equ_fuzzy}) becomes,
  \begin{equation}
    \mathbf{f}=\mathbf{F}\,\mathbf{a}
    \label{equ_fuzzy2}
  \end{equation}
where $\mathbf{F}$ is an $[N\times N]$ matrix containing the
elements of the Gaussian function.

\emph{How do we determine the optimum $\omega$ given the observed
  data, kernel and statistical noise?}

The tools for addressing this question have been presented. That
is, we employ Bayes' theorem to determine the \emph{a posteriori}
probability distribution for $\omega$ conditional on the observed
line profile. The $\omega$ which maximises the resulting  \emph{a
posteriori} probability distribution becomes the optimum fuzzy
pixel width, $\hat{\omega}$. At a practical level, we replace the
equations where $\mathbf{P}$ appears with $\mathbf{a}$, and the
kernel $\mathbf{K}$ is replaced by
\begin{equation}
  \mathbf{G}=\mathbf{R}\,\mathbf{F} \label{equ_newkernel}
\end{equation}
where $\mathbf{G}\equiv\mathbf{G}(\omega)$.

Applying Bayesian theory, the distribution for $\omega$ can be
determined by integrating out $\mathbf{a}$ and $\alpha$,
\begin{subequations}
  \begin{eqnarray}
    \Pr(\omega|\,\mathbf{g},\,\mathbf{m},\,\mathbf{\sigma},\,\mathcal{I}) &=&  \int
\mathcal{D}\mathbf{a} \, \int \mathrm{d}\alpha
\Pr(\mathbf{a},\,\alpha,\,\omega|\,\mathbf{g},\,\mathbf{m},\,\mathbf{\sigma},\,
\mathcal{I})\label{equ_jpdomega}\\
&=& \int\mathcal{D}\mathbf{a} \, \int \mathrm{d}\alpha
\,\Pr(\alpha| \, \mathcal{I})
\,\Pr(\omega| \, \mathcal{I}) \nonumber\\
&\times&
\Pr(\mathbf{a}|\,\mathbf{g},\,\mathbf{m},\,\mathbf{\sigma},\,\alpha,\,\omega,\,
\mathcal{I}).\label{equ_jpdomega2}
  \end{eqnarray}
Following the same steps as in (\ref{equ_subequ1}), we have
\begin{eqnarray}
\Pr(\omega|\,\mathbf{g},\,\mathbf{m},\,\sigma,\,\mathcal{I})
&\approx&
\frac{\Pr(\alpha|\,\mathcal{I})\,\Pr(\omega|\,\mathcal{I})}{\Pr(\mathbf{g}|\,\mathbf{m},\,\sigma,\,\mathcal{I})}\,\frac{1}{Z_{S}(\alpha)\, Z_{L}(\sigma)} \nonumber\\
&\times& \frac{(2\pi)^{\frac{N}{2}}\,e^{Q(\alpha,\,\omega)}
  }{\sqrt{\det \nabla \nabla Q(\alpha,\,\omega)}}. \label{equ_jpdomega3}
  \end{eqnarray}\label{equ_subequ2}
\end{subequations}
where $Q(\mathbf{a},\, \alpha, \, \omega) =\alpha\,S(\mathbf{a}) -
L(\mathbf{a},\,\omega)$ for the unknown terms $\mathbf{a}$,
$\alpha$ and $\omega$; and
$\nabla \nabla Q(\alpha,\, \omega) \equiv \nabla_{\mathbf{a}}\nabla_{\mathbf{a}}Q(\alpha,\, \omega)$. 

Error bars can also be attributed to $\mathbf{a}$ and
$\mathbf{f}$. Using the results discussed in
\S~\ref{sec_errorbars}, the covariance matrix for $\mathbf{a}$,
$\nabla_{\mathbf{a}} \nabla_{\mathbf{a}} Q$ can be determined. The
corresponding covariance matrix for $\mathbf{f}$ can be determined
from $\nabla_{\mathbf{f}} \nabla_{\mathbf{f}}
Q=\mathbf{F}\,\nabla_{\mathbf{a}} \nabla_{\mathbf{a}}
Q\,\mathbf{F}^{T}$. On applying (\ref{equ_var}) the error bars for
$\mathbf{f}$ can be determined.

 Traditionally this problem has been solved by applying classical
 techniques, such as the \citeasnoun{stokes48} method. In order
to overcome the numerical instability of the Stokes method,
methods such \emph{direct convolution}
 \cite{howard89a,howard89b}  and profile fitting methods, such as
 the Voigt function
 \cite{wu98a,balzar93a,langford92,keijser83,keijser82} have been developed. These approaches
assume an analytical profile function for the specimen profile;
the convolution product between the instrument and specimen
profile is refined, by updating the parameters that define the
specimen profile, until the error between the calculated and
observed data is minimized. These methods are a means to an end.
There is often no physical basis for choosing a particular profile
function, except that it results in a minimized error
\cite{armstrong99}. However, the Bayesian/fuzzy pixel/MaxEnt approach determines the \emph{maximally uncommitted}
solution or the solution with the least assumptions \cite{wu97},
given all the available data and information.

\section{Generating \& analyzing simulated $\mbox{CeO}_{2}$ data}\label{sec_mockdata}

\subsection{Generating the simulated data}\label{sec_genmock}
\paragraph{Particle-size distribution, $P(D)$.}
In order to test the Bayesian/MaxEnt method, simulated data for
the $2 0 0$ and $4 0 0$ line profiles from $\mbox{CeO}_{2}$ were
generated. The crystallites were assumed to be spherical in shape
with a log-normal crystallite-size distribution,
\begin{subequations}
  \begin{equation}
P(D)=\frac{1}{\sqrt{2\pi\,D^{2}\,\ln^{2}\sigma_{0}}}\,\exp\left[-\frac{1}{2}
\left(\frac{\ln (D/D_{0})}{\ln\sigma_{0}}\right)^{2}\right]
    \label{equ_lognorm}
\end{equation}
where $D_{0}$ is the median and $\sigma_{0}^{2}$ is the log-normal
variance. The average diameter, $\langle D \rangle$, and variance, $\sigma_{\langle D
\rangle}^{2}$, of the distribution are related to these
quantities by
\begin{equation}
    \langle D \rangle= D_{0}\,e^{\ln^{2}\sigma_{0}/2} \label{equ_logave}
\end{equation}
and
\begin{equation}
   \sigma_{\langle D \rangle}^{2} =
   D_{0}^{2}\,e^{\ln^{2} \sigma_{0}}\left(e^{\ln^{2}\sigma_{0}} - 1\right).
\label{equ_logvar}
\end{equation}
\label{equ_lognormal}
\end{subequations}

The log-normal parameters used were $D_{0}=13.03\,\mbox{nm}$ and
 $\sigma_{0}^{2}=2.89$. Using
(\ref{equ_logave}) and (\ref{equ_logvar}), the average diameter
and variance were determined to be, $\langle D \rangle
=15.00\,\mbox{nm}$ and $\sigma_{\langle D
\rangle}^{2}=73.17\,\mbox{nm}^{2}$, respectively. Using the
results from \citeasnoun{krill98}(see Equations 6-8, p625), the
corresponding area- and volume-weighted sizes were determined.

The area- and volume-weighted diameters for spheres are related to
the sizes \cite{langford00} by
\begin{subequations}
  \begin{equation}
    \langle D \rangle_{a}=\frac{3}{2}\langle t \rangle_{a}
    \label{equ_areadiam}
  \end{equation}
and
  \begin{equation}
    \langle D \rangle_{v}=\frac{4}{3}\langle t \rangle_{v}
    \label{equ_voldiam}.
  \end{equation}
\label{equ_diam}
\end{subequations}
The area- and volume-weighted sizes, $\langle t \rangle_{a}$ and
$\langle t \rangle_{v}$, can be determined from the specimen profile,
$f$, and Fourier coefficients, $A(t)$, by using \cite{warren69}
\begin{equation}
  \langle t \rangle_{a}^{-1} = -\left. \frac{\mathrm{d}A(t)}{\mathrm{d}t}
   \right |_{t\rightarrow 0}
  \label{equ_areat}.
\end{equation}
The volume-weighted size is inversely related to the integral
breadth and can be determined either directly from the specimen
profile, $f$, or from its Fourier coefficients, $A(t)$,
\begin{subequations}
 \begin{eqnarray}
  \beta &=& \int_{-\infty}^{\infty}\, f(s)\,\mathrm{d}s /f_{max}
  \label{equ_int1} \\
        &=& \left[ 2 \int_{0}^{\infty}\, A(t)\,\mathrm{d}t
        \right]^{-1} \label{equ_int2}\\
        &=& \langle t \rangle_{v}^{-1} \label{equ_volt},
 \end{eqnarray}
\label{equ_intbreadth}
\end{subequations}
where $\beta$ is in reciprocal space units.

Using (\ref{equ_areat}) and (\ref{equ_diam}), the area-weighted
size and diameter were determined  as $\langle t\rangle_{a}=17.56
\,\mbox{nm}$ and $\langle D \rangle_{a}=26.34\,\mbox{nm}$,
respectively. Using (\ref{equ_int1}) and (\ref{equ_diam}), the
volume-weighted size and diameter were determined
as $\langle t \rangle_{v}=26.18\,\mbox{nm}$ and $\langle D \rangle_{v}=34.91
\,\mbox{nm}$, respectively. These settings are considered as the
theoretical values for the simulated data. The Bayesian/fuzzy
pixel/MaxEnt results are compared with the theoretical sizes and
percentage differences are determined.

\paragraph{Line profiles, $f(2\theta)$ \& $k(2\theta)$.}
Using the parameters for the size distribution, the specimen profile
for spherical crystallite, $f(2\theta)$, was modelled over the
range $(2\theta_{0}
 \pm10)\,^{\circ}2\theta$ at a step size of $0.01 \,^{\circ}2\theta$ (see
(\ref{equ_sprofile4})). The simulation of the specimen profile
over this range minimized any artifacts in the Fourier
coefficients. The instrument profile, $k(2\theta)$, was modelled
on the diffractometer parameters and $\mbox{LaB}_{6}$
line-position standard reference material, as discussed in
\S~\ref{sec_exp}. The split-Pearson VII function for the 200 line
consisted of the following parameters:
$\mbox{FWHM}_{low}=0.030\,^{\circ}2\theta$,
$\mbox{FWHM}_{high}=0.027\,^{\circ}2\theta$ and
$m_{exp,low}=6.928$, $m_{exp,high}=11.324$, where $m_{exp}$ are
the  split-Pearson exponents. The `low' and `high' subscripts are
with respect to the Bragg positions, $2\theta_{0}$ (see
Fig.~\ref{fig_exp_lab6}).

\paragraph{Generating $g(2\theta)$.}
The  observed line profiles, $g(2\theta)$, for the $2 0 0$ and $4 0
0$ lines consisted of the convolution of the specimen line
profile, $f(2\theta)$, with the instrument line profile,
$k(2\theta)$, Poisson noise, and a linear background level,
$b(2\theta)$. Statistical noise was also imparted onto the
background  before adding it to the convoluted product. This is
expressed by (\ref{equ_obs}).

\begin{figure}[htb!]
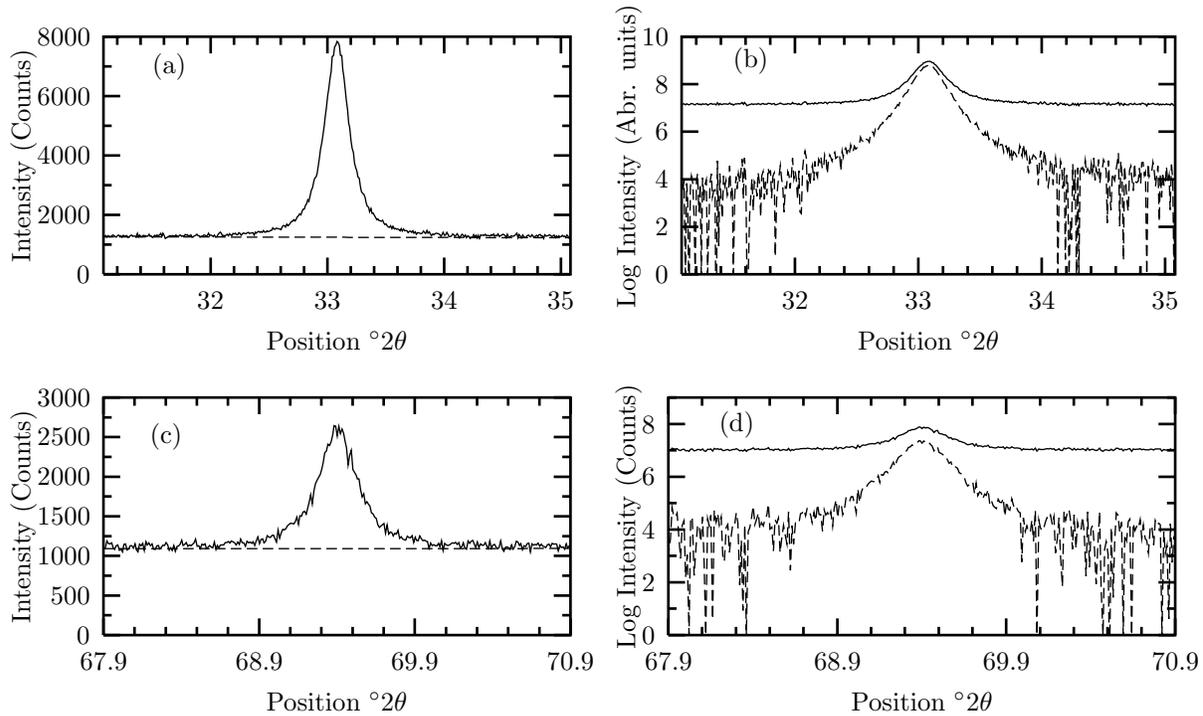

\begin{tabular}{cc}
{\footnotesize\input{size_fig_obs_sim200.tex}} &
{\footnotesize\input{size_fig_logobs_sim200.tex}}\\
{\footnotesize\input{size_fig_obs_sim400.tex}} &
{\footnotesize\input{size_fig_logobs_sim400.tex}}
\end{tabular}
\vspace{-5mm} \caption{\footnotesize Simulated observed $2 0 0$
and $4 0 0$ CeO$_{2}$ profiles, $g(2\theta)$. (a) The $2 0 0$
profile (solid line) and estimated background level (dashed line)
over $(2\theta_{0}\pm 2)\,^{\circ}2\theta$, the range over which
the analysis was carried out. (b) Logarithm of the  $2 0 0$
profile before (solid line) and after (dashed line) the background
estimation. (c \& d) Plots corresponding to (a) and (b),
respectively, for the $4 0 0$ profile over $(2\theta_{0}\pm
1.5)\,^{\circ}2\theta$ \label{fig_obs_ceo2}}
\end{figure}

The generation of  $g(2\theta)$ was carried out over
$2\theta_{0}\pm 10\,^{\circ}2\theta$ in order to minimize any
truncation errors.  The maximum peak height for the $2 0 0$ line
profile was set to 6500 counts (without background level and
noise or a total of 7835 counts including background level and
noise) and the peak-to-background ratio, $R_{pb}$, was set to 6.0.
The corresponding  percentage error in the peak maximum was
determined using
\begin{equation}
  \sigma_{peak} =
  \frac{1}{(R_{pb}-1)}\,\left[\frac{R_{pb}(R_{pb}+1)}{
  I_{max,bg}}\right]^{\frac{1}{2}}
  \, \times \,100 \%
  \label{equ_errorcounts}
\end{equation}
where $I_{max,bg}$ is the maximum number of counts, including
background level. Simulated  $g(2\theta)$ for the $2 0 0$ and $4 0
0$ line profiles are shown in Fig.~\ref{fig_obs_ceo2}. The
uncertainty for the $2 0 0$  line was  1.5\% in the peak height.
Similarly, for the $4 0 0$  line the maximum peak height was set
to 1500 counts (2646 counts including background level and noise),
the average peak-to-background ratio was set to 2.4; and the
estimated statistical uncertainty in the peak height was found to
be 4.0\%.

In order to simulate realistic conditions, the Bayesian/MaxEnt
analysis of the  $g(2\theta)$ was carried out in a truncated
region $(2\theta_{0}\pm 2)\,^{\circ}2\theta$ for the $2 0 0$
and $(2\theta_{0}\pm 1.5)\,^{\circ}2\theta$ for the $4 0 0$ line
profiles. In the analysis, the background level was assumed to be
unknown and was approximated by a linear function over this
region. This was achieved by examining the Fourier coefficients of
$g(2\theta)$ as the level was raised/lowered until distortions
(i.e. `hook effect' etc.) were removed. Fig.~\ref{fig_obs_ceo2}
shows the simulated $g(2\theta)$ before and after the background level
estimation for the $2 0 0$ and $4 0 0$ line profiles.

\paragraph{Generating the kernels, $\mathbf{R}$, $\mathbf{I}_{p}$ \&
  $\mathbf{K}$ }\label{sec_kernels}
The numerical evaluation of the instrument kernel $\mathbf{R}$,
line profile kernel $\mathbf{I}_{p}$, and scattering kernel
$\mathbf{K}$, are an important aspect in the application of the
Bayesian/MaxEnt method. The evaluation of the fuzzy pixel kernel,
$\mathbf{F}$, is also important in the implementation of the fuzzy
pixel/MaxEnt method in determining the specimen profile, $f$. This
section expands on \S~\ref{sec_observations}.

The advantage of the \citename{bryan90a} algorithm
\cite{bryan90a} and the Bayesian/MaxEnt
algorithm  is that the search direction (or subspace) is defined
by the singular value decomposition (SVD) of the scattering
kernel, $\mathbf{K}$. This approach is numerically efficient (in
that it reduces the number of floating point operations) and also
numerically stable, since it does not utilize the full
column-space of the kernels. As was pointed out in
\S~\ref{sec_observations}, the vector-space spanned by the column
vectors of $\mathbf{K}$ may be all (or nearly all) linearly
dependent, causing it to be ill-conditioned. The ill-conditioned
characteristics are overcome by the SVD of $\mathbf{K}$,
$\mathbf{V}\,\mathbf{\Sigma}\,\mathbf{U}^{T}$, where the `singular
space' spanned by the column vectors of $\mathbf{U}$ is used to
define the subspace in which the size distribution can be
determined.

The instrument kernel, $\mathbf{R}$, is an $[M\times N']$ matrix.
The elements of this matrix can be determined by
$R_{ij'}=k(2\theta_{i}-2\theta_{j'})$, where $M \geq N'$. This
matrix can be mapped into reciprocal-space, s, by multiplying each
column of $\mathbf{R}$ by $\mathrm{d}(2\theta)/\mathrm{d}s
=\lambda/\cos \theta_{j'}$.

The line profile kernel expresses (\ref{equ_intprofile}) as an
$[N' \times N]$ matrix, $\mathbf{I}_{p} \equiv [I_{p\,j'j}$],
consisting of the line profile from a specific common volume (i.e.
shape) function. The formalism presented here is completely
general and any shape function can be used where appropriate. In
this study, we have employed the common-volume function for
spherical crystallites (see \ref{equ_sphereexample}),
\begin{equation}
  I_{p\,j'j}=\left\{ \begin{array}{ll}
      \frac{1}{16 \pi^{3} s_{j'}^{\prime 4}} + \frac{D_{j}^{2}}{8 \pi
      s_{j'}^{\prime 2}} - \frac{\cos(2\pi s_{j'}' D_{j})}{16 \pi^{3}
      s_{j'}^{\prime 4}} - \frac{D_{j}\, \sin(2\pi s_{j'}^{\prime}
      D_{j})}{8 \pi^{2} s_{j'}^{\prime 3}} & \mbox{for $ s_{j'}' \neq 0$} \\
            \frac{\pi D_{j}^{4}}{8} & \mbox{for $ s_{j'}' =0$}
                    \end{array}
              \right. \label{equ_lineprofilesphere}
\end{equation}
where the second term in (\ref{equ_lineprofilesphere}) ensures
that the line profile from a single spherical crystallite is
finite for $s=0$.

The evaluation of the scattering kernel, $\mathbf{K}$, is the
matrix product of the instrument kernel (mapped into $s$-space),
$\mathbf{R}$, and the line profile kernel,
(\ref{equ_lineprofilesphere}). Using (\ref{equ_kernel})
\begin{subequations}
  \begin{eqnarray}
    K(s_{i},\,D_{j}) &=& \delta D \, \delta s'\,
    \sum_{j'}\,k(s_{i}-s_{j'}')\,I_{p}(s_{j'}',\,D_{j})\label{equ_scatkernela} \\
    K_{ij} &=& \delta D \, \delta s'\, \sum_{j'}\,R_{ij'}\,I_{p\,j'j}
 \label{equ_scatkernelb} \\
   \mathbf{K} &=&  \delta D \, \delta s'\,\mathbf{R}\, \mathbf{I}_{p}\label{equ_scatkernelc}
  \end{eqnarray}
\end{subequations}
where $\mathbf{R}$ has been mapped into $s$-space, $\delta s'$ is
the step size in $s'$-space and approximates the integration in
(\ref{equ_kernel}), while $\delta D$ is the step size in $D$-space
and approximates the integration in (\ref{equ_single}). Care must
be taken in selecting $\delta D$ to avoid the under-sampling of
(\ref{equ_lineprofilesphere}).

\subsection{Applying the fuzzy pixel/MaxEnt method for
  $f(2\theta)$}\label{sec_mockfuzzy}

This approach involves determining the specimen profile from the simulated data. It is equivalent to solving the deconvolution problem,
(\ref{equ_obs}), and is an important first step in assessing the
nature of the specimen broadening. In the past, we have applied
the \citeasnoun{skilling84} algorithm with global smoothing
\citeaffixed{sivia96}{see}, which we refer to here as the `old'
MaxEnt method. However, in this section we apply the Fuzzy
Pixel/MaxEnt method discussed in \S~\ref{sec_fuzzy}, to determine
$f(2\theta)$. The results are also compared with those from the
`old' MaxEnt method, and their reliability in reproducing the
log-normal parameters for the crystallite-size distribution (set
in \S~\ref{sec_genmock}) are assessed.

\begin{figure}[thb!]
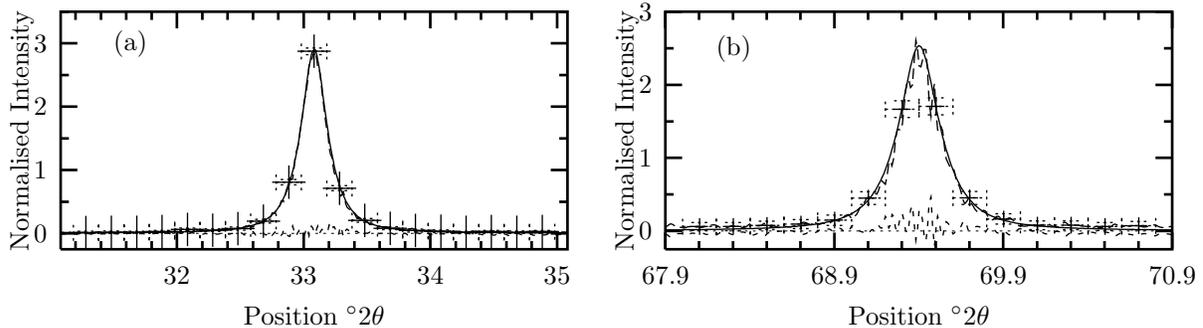

\begin{tabular}{cc}
{\footnotesize\input{size_fig_oldfsol_sim200.tex}} &
{\footnotesize\input{size_fig_oldfsol_sim400.tex}}
\end{tabular}
\caption{\footnotesize `Old' MaxEnt specimen profiles for the $2 0
0$ and $4 0 0$ line profiles. (a)  The theoretical $2 0 0$
specimen profile (solid line), `old' MaxEnt specimen profile (long
dashed line + error  bars) and the residuals (short dashed line).
(b) Corresponding results for the $4 0 0$ line profile, as shown
in (a). The horizontal error bars in (a) \& (b) represent
non-overlapping region  of interest, while the vertical error bars
represent the uncertainty in the averaged integrated flux over the
region of interest. \label{fig_oldsol}}
\end{figure}

 The specimen line profiles from the `old' MaxEnt approach are given
in Fig.~\ref{fig_oldsol}. These results were compared with the
theoretical specimen profiles by evaluating the $R_{f}$ and
$R_{w}$ values. A summary of these and subsequent analyses is
given in Table~\ref{tab_memresults}.

\begin{figure}[thb!]
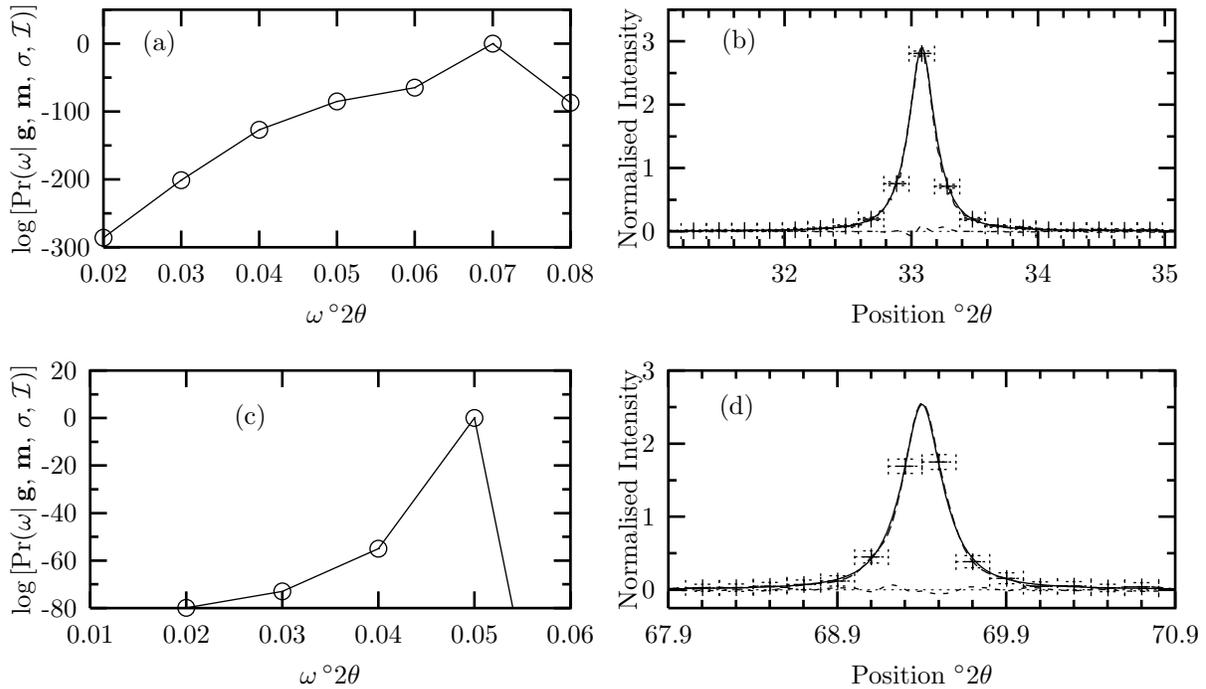

\begin{tabular}{cc}
{\footnotesize\input{size_fig_fuzzyw_sim200.tex}} &
{\footnotesize\input{size_fig_fuzzyfsol_sim200.tex}}\\
{\footnotesize\input{size_fig_fuzzyw_sim400.tex}} &
{\footnotesize\input{size_fig_fuzzyfsol_sim400.tex}}
\end{tabular}
\caption{\footnotesize The fuzzy pixel distribution and MaxEnt
  solutions for the $2 0 0$ and $4 0 0$ line profiles. (a) The $\log
  \Pr(\omega|\,\mathbf{g},\,\mathbf{m},\,\sigma,\,\cal{I})$
  distribution used to determine the optimum fuzzy pixel width,
  $\hat{\omega} \approx 0.07\,^{\circ}2\theta$ for the $2 0 0$ specimen proifle. (b) Theoretical specimen profile (solid line),
  fuzzy pixel/MaxEnt specimen profile (long dashed line + error
  bars) and the residuals (short dashed line) for the $2 0 0$ line
  profile. (c) The $\log
  \Pr(\omega|\,\mathbf{g},\,\mathbf{m},\,\sigma,\,\cal{I})$
  distribution used to determine the optimum fuzzy pixel width,
  $\hat{\omega} \approx 0.05\,^{\circ}2\theta$ for the $4 0 0$
  specimen proifle. (d) Theoretical specimen
  profile (solid line), fuzzy pixel/MaxEnt specimen profile (long
  dashed line + error bars) and the residuals (short dashed line) for
  the $4 0 0$ line profile.
  The horizontal error bars in
  (a) \& (c) represent the non-overlapping region of interest, while the vertical
  error bar represents the uncertainty in the averaged integrated flux
  over the region of interest.
  \label{fig_fuzzsol}}
\end{figure}

The `old' MaxEnt method is not based on a Bayesian formalism
\citeaffixed{gull89,skilling89a}{see} and spurious oscillations
can appear in the solution specimen profile. This second point
becomes important in analyzing high angle/low intensity profiles.
This is further illustrated by inspecting the residuals in
Fig.~\ref{fig_oldsol}(b), where the amplitude of the residuals is
large in comparison with the normalized peak height. We contrast
the results in Fig.~\ref{fig_oldsol}  with the fuzzy pixel/MaxEnt
method discussed in \S~\ref{sec_fuzzy}. Using this theory, the
fuzzy pixel distribution specimen profiles are shown in
Fig.~\ref{fig_fuzzsol}. The fuzzy pixel distribution determines
the optimum fuzzy pixel width, $\omega$ (see (\ref{equ_subequ2})).
For the $2 0 0$ line, the optimum value was found to be
$\hat{\omega} \approx 0.07\,^{\circ}2\theta$ and for the $4 0 0$
line, $\hat{\omega} \approx 0.05\,^{\circ}2\theta$. This defines
the correlation-length scale of the noise in the simulated data
and essentially filters out the noise effects. It is evident from
the residuals of the multiple orders that smoothing of the
specimen profile has been achieved using this approach.

Using the line profiles determined and assuming a spherical
crystallite shape, the parameters of the underlying log-normal
size distribution can be reproduced by following the approach of
\citeasnoun{krill98}. These results are shown in
Table~\ref{tab_memresults}. The analysis has produced mixed
results, due to the stringent but realistic conditions imposed on
the background estimation. Comparing the $2 0 0$ line profile
results for the `old' MaxEnt and fuzzy pixels methods, there is a
noticeable improvement in the latter results over the former. This
is not only seen in an improved $R_{f}$ value, but also in the
reproduced log-normal parameters. In the case of the $4 0 0$ line,
we notice that the $R_{f}$ value has improved by a factor of $\sim
3$ and the volume-weighted size by a factor of $\sim 1.5$ for the
fuzzy pixel/MaxEnt approach. However, the area-weighted size for
the $4 0 0$ line profile has not improved. As a consequence, when
the underlying log-normal parameters are determined from the area-
and volume-weighted sizes no improvements are gained.
\begin{table}[htb!]
\footnotesize
\begin{center}
\begin{tabular}{lcc|cc}\hline \hline
 & \multicolumn{2}{c|}{'old' MaxEnt} & \multicolumn{2}{c}{Fuzzy Pixel/MaxEnt}\\
\cline{2-5}
  Results      & $2 0 0$ & $4 0 0$   &   $2 0 0$     &  $4 0 0$ \\
\hline \hline
  $R_{f}\, (\%)$
  &   4.2       &   10.9    &  2.7         &   3.1        \\
  $R_{w}\, (\%)$
  &   2.9       &   3.7     &  3.0         &   3.7         \\
  $\langle t \rangle_{a}\,\mbox{(nm)}$
  & $\begin{array}{c}19.9 \pm 0.1 \,(13.3\%)\end{array}$ &
    $\begin{array}{c}20.4 \pm 0.1 \, (16.0\%)\end{array}$ &
    $\begin{array}{c}17.89\pm 0.07 \,  (1.9\%)\end{array}$ &
    $\begin{array}{c}20.5 \pm 0.1 \, (16.5\%) \end{array}$\\
  $\langle D \rangle_{a}\,\mbox{(nm)}$
  & $29.8 \pm 0.2$ & $30.5\pm 0.2$  & $26.8 \pm 0.1$   & $30.7 \pm 0.2$\\
  $\langle t \rangle_{v}\,\mbox{(nm)}$
  & $\begin{array}{c}26.63\pm 0.07 \,(1.7\%)\end{array}$  &
   $\begin{array}{c}28.0\pm 0.2 \, (7.1\%)\end{array}$ &
   $\begin{array}{c}25.86 \pm 0.04 \, (1.2\%) \end{array}$ &
   $\begin{array}{c} 27.4 \pm 0.2 \, (4.8\%)\end{array}$ \\
  $\langle D \rangle_{v}\,\mbox{(nm)}$ &
  $35.51\pm 0.09$ & $37.4\pm 0.2$ & $34.48\pm 0.05$   & $36.6\pm 0.3$ \\
  $D_{0}\,\mbox{(nm)}$
  & $\begin{array}{c}19.3\pm 0.4 \,(48.1\%)\end{array}$ &
    $\begin{array}{c}18.4 \pm 0.5\, (41.5\%)\end{array}$ &
    $\begin{array}{c}14.3\pm 0.2 \, (10.1\%) \end{array}$ &
    $\begin{array}{c}19.8 \pm 0.6 \, (52.0\%) \end{array}$ \\
  $\sigma_{0}$
& $\begin{array}{c}1.52 \pm 0.01 \, (10.7\%)\end{array}$ &
  $\begin{array}{c}1.57 \pm 0.02 \, (7.8\%) \end{array}$ &
  $\begin{array}{c}1.650\pm 0.007 \, (3.0\%)\end{array}$  &
  $\begin{array}{c}1.52 \pm 0.02\, (10.6\%) \end{array}$ \\
$\langle  D \rangle \,\mbox{(nm)}$
  & $\begin{array}{c}21.06\pm 0.43 \, (40.4\%)\end{array}$ &
    $\begin{array}{c}20.4\pm 0.5 \, (36.0\%) \end{array}$ &
    $\begin{array}{c}16.3 \pm 0.2 \, (8.4\%) \end{array}$ &
    $\begin{array}{c}2.6 \pm 0.6 \, (44.1\%)\end{array}$  \\
  $ \sigma_{\langle D \rangle}^{2} \,(\mbox{nm}^{2})$
  & $\begin{array}{c}84 \pm 5 \, (15.3\%)\end{array}$ &
   $\begin{array}{c} 93\pm 7 \, (27.2\%)\end{array}$ &
   $\begin{array}{c}75 \pm 3 \, (2.9\%)\end{array}$  &
$\begin{array}{c}89 \pm 8 \, (22.4\%) \end{array}$ \\
\hline \hline
\end{tabular}
\caption{\footnotesize Area- and volume-weighted sizes for the $2
0 0$ and $4 0 0$ specimen line profiles ($f$) from the `old'
MaxEnt and fuzzy pixel/MaxEnt methods. $\langle D \rangle_{a}$ and
$\langle D \rangle_{v}$ values were determined using
(\ref{equ_diam}).  The $D_{0}$, $\sigma_{0}$, $\langle D \rangle$
and $\sigma_{\langle D \rangle}^{2}$ values were determined from
(\ref{equ_lognormal}). The percentage differences between the
calculated and theoretical values are given in parentheses.
\label{tab_memresults}}
\end{center}
\end{table}

These results for the $4 0 0$ line profile can be explained by the
low peak-to-background ratio, statistical uncertainty, and the
presence of systematic errors arising from the background
estimation. The peak-to-background ratio for the $4 0 0$ line
profile is 2.4. This low value results in an increased uncertainty
in the estimated background level. From (\ref{equ_errorcounts}),
we notice that as the peak-to-background ratio increases, the peak
height uncertainty decreases and the dominant source of
uncertainty becomes the statistical noise.\footnote{Taking
(\ref{equ_errorcounts}), we see that in the limit of $R_{pb}
\rightarrow 1$, $\sigma_{p}\rightarrow \infty$. On the other hand,
in the limit of $R_{pb} \rightarrow \infty$,
$\sigma_{p}\rightarrow 1/\sqrt{I_{max,bg}}$. For example, with $R
\sim 15$, $\sigma_{p} \sim 1.2/\sqrt{I_{max,bg}}$.} The variance
of the observed profile is determined by two components: the
Poisson counting statistics, which can be approximated as
$\sqrt{\mbox{g}}$, for $\mbox{g}>>10\,\mbox{counts}$, and the
estimated background level, $b_{est}$; it can be expressed as
$\sigma^{2}=g+ b_{est}$. The presence of statistical uncertainty
and the low peak-to-background ratio introduces uncertainties to
the slope and intercept of the estimated background level. In
turn, this introduces systematic errors to the Fourier
coefficients of $f$ \cite{armstrong99}. Although the fuzzy/pixel
method has been successful in improving the quality of the line
profile (which amounts to reducing the statistical error in the
solution line profile), the systematic errors have propagated to
the Fourier coefficients of the specimen profile and in turn
to the area-weighted size. Additional calculations and applying
the above analysis to simulated data with zero background (i.e.
only Poisson noise) show percentage differences between the
calculated and theoretical results of $\lesssim 5\%$ for both the
$2 0 0$ and $4 0 0$ fuzzy pixel/MaxEnt specimen profiles. This
highlights the difficulty of analyzing high-angle/weak line
profiles, which clearly requires a good understanding of the
background level in order to reduce the influence of systematic
errors.

The application of the fuzzy pixel/MaxEnt method for determining
$f(2\theta)$ enables the specimen broadening to be assessed. This
is important in the application of methods such as those of 
Warren-Averbach and Williamson-Hall. Furthermore, the
analysis discussed here can be used as the \emph{a priori}
information of the Bayesian/MaxEnt analysis. The fuzzy
pixel/MaxEnt approach overcomes the difficulties in commonly used
deconvolution techniques \citeaffixed{armstrong98a}{see} and
resolves the `ringing effects' in \citeasnoun{armstrong99b}.

\subsection{Bayesian/MaxEnt method for $P(D)$ using different
$m(D)$}\label{sec_mockanal}

The next stage in the analysis of the simulated data is applying
the Bayesian/MaxEnt method to determine the particle distribution,
$P(D)$. In addition, two different approaches for determining a
model, $m(D)$, were explored and their effects on $P(D)$ were
quantified. The two approaches were (i) uniform model over
$D\in[0,\,60]\,\mbox{nm}$ and (ii) `low resolution' approach
\cite{armstrong99} using the log-normal distribution parameters
determined in \S~\ref{sec_mockfuzzy} as the prior.

\begin{figure}[thb!]
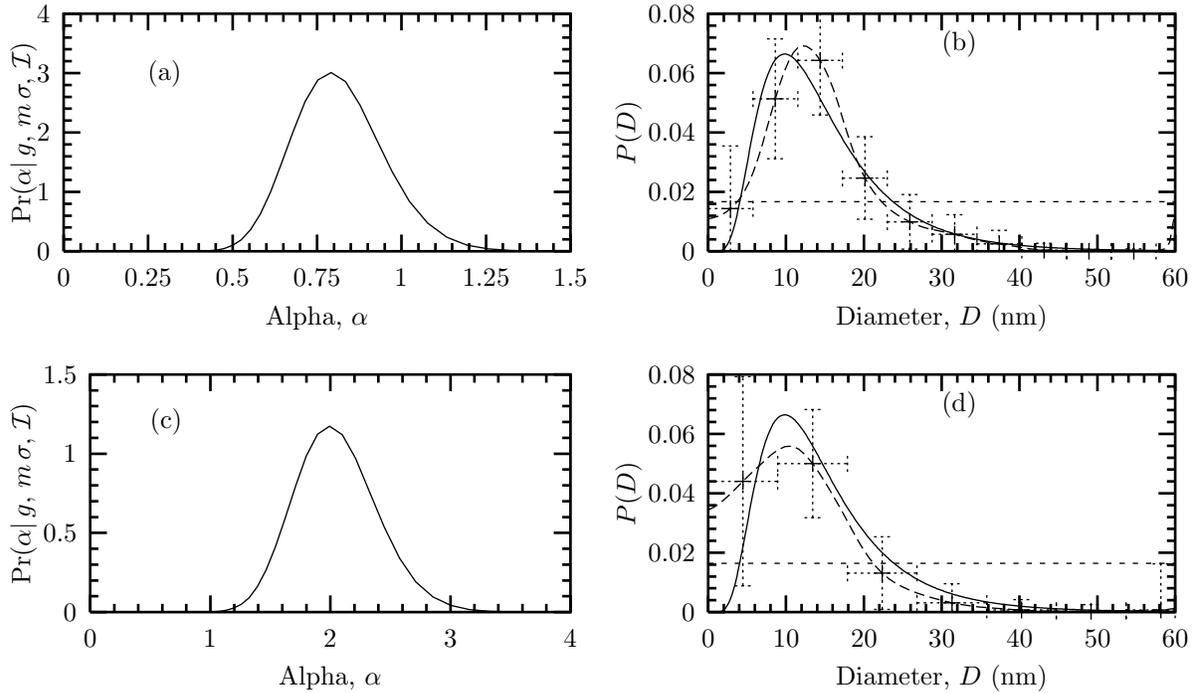

\begin{tabular}{cc}
{\footnotesize\input{size_fig_alpha_sim200uni.tex}} &
{\footnotesize\input{size_fig_psol_sim200uni.tex}}\\
{\footnotesize\input{size_fig_alpha_sim400uni.tex}} &
{\footnotesize\input{size_fig_psol_sim400uni.tex}}
\end{tabular}
\caption{\footnotesize   Bayesian/MaxEnt crystallite size
distributions using a uniform \emph{a priori} model. (a) The
$\Pr(\alpha|\, \mathbf{g},\,\mathbf{m},\,\sigma,\,\cal{I})$
distribution, (\ref{equ_subequ1}), used to average over the set of
solutions, $\{\mathbf{P}(\alpha)\}$. (b) The theoretical
crystallite size distribution (solid line), Bayesian/MaxEnt size
distribution (long dashed line + error bars), and the uniform
\emph{a priori} model (short dashed line). (c) \& (d) The
corresponding $\Pr(\alpha|\,
\mathbf{g},\,\mathbf{m},\,\sigma,\,\cal{I})$ distribution and
Bayesian/MaxEnt size distribution for the $4 0 0$ line profile.
\label{fig_unibayessol}}
\end{figure}

\paragraph{Uniform model}
The Fourier coefficients $A(t)$ of the fuzzy pixel/MaxEnt specimen
profiles (not shown here) suggest the maximum size of the
crystallites is $\sim 60\,\mbox{nm}$, since $A(t)\sim 0$ at this
length. Using this information, a uniform distribution was defined
over $D\in[0,\,60]\,\mbox{nm}$. The corresponding Bayesian/MaxEnt
results are shown in Fig.~\ref{fig_unibayessol}. The posterior
distribution for $\alpha$ is shown in
Figs.~\ref{fig_unibayessol}(a) \& (c) for the $2 0 0$ and $4 0 0$
profiles, respectively. This distribution was used to average over
the set of solutions $\{\mathbf{P}\}$ for each case. The
Bayesian/MaxEnt results are given in Fig.~\ref{fig_unibayessol}(b)
\& (d) for the $2 0 0$ and $4 0 0$ profiles respectively.

Using a uniform model, the Bayesian/MaxEnt size distributions
where compared with the theoretical size distribution, $P(D)$. The
Bayesian/MaxEnt results share `global' features with
the theoretical size distributions. However, `local' features are
poorly defined, especially in the region of $0\lesssim D \lesssim
10\, \mbox{nm}$. This is a direct consequence of the uniform
model and the lack of relevant information in the data; that is, it assigns an equal weight to all sizes over $D$.
The vertical error bars in both cases correctly represent the
misfitting between the theoretical and Bayesian/MaxEnt size
distributions; additionally, their magnitude also signifies that a
uniform model transfers little or no useful information. This can
also be seen in the parameters for the Bayesian/MaxEnt
distribution compared with their theoretical values in
Table~\ref{tab_unip}. In determining the log-normal parameters
from the Bayesian/MaxEnt $P(D)$, the fitted distribution produce
reasonable results. This suggests that, although the \emph{a
proiri} model is uniform, the Bayesian/MaxEnt method can `extract'
some information concerning the underlying distribution from the
simulated data.

\begin{table}[thb!]
\footnotesize
\begin{center}
\begin{tabular}{lcc|cc}\hline \hline
 & \multicolumn{2}{c|}{Uniform model}& \multicolumn{2}{c}{`Low'
   Res. Model} \\ \cline{2-5}
Results
& $2 0 0$ & $4 0 0$ & $2 0 0$ & $4 0 0$ \\ \hline
$R_{f}\,(\%)$
& 23.0  & 40.0  & 22.2   & 19.1 \\
$D_{0} \, (\mbox{nm})$
& $\begin{array}{c}13.9 \pm 0.3 \, (6.5\%) \end{array}$ &
   $\begin{array}{c}11.9 \pm 0.9 \, (8.8\%) \end{array}$ &
   $\begin{array}{c}14.8 \pm 0.2 \, (13.4\%)\end{array}$ &
   $\begin{array}{c}12.5 \pm 0.2 \, (4.4\%) \end{array}$ \\
$\sigma_{0}$
& $\begin{array}{c}1.589\pm 0.003 \, (6.5\%) \end{array}$ &
$\begin{array}{c} 2.14 \pm 0.03, (25.8\%) \end{array}$ &
$\begin{array}{c}1.612 \pm 0.002\, (5.1\%) \end{array}$ &
   $\begin{array}{c}1.544 \pm 0.002 \, (9.2\%) \end{array}$ \\
$\langle D \rangle \,\mbox{(nm)}$ &
$\begin{array}{c}15.5 \pm 0.3 \, (3.0\%)\end{array}$  &
$\begin{array}{c} 15 \pm 2 \, (5.8\%) \end{array}$ &
$\begin{array}{c}16.6\pm 0.2 \,  (10.4\%)\end{array}$ &
$\begin{array}{c}13.7 \pm  0.2 \, (8.7\%) \end{array}$\\
$ \sigma_{\langle D \rangle}^{2} \,(\mbox{nm}^{2})$
& $\begin{array}{c}57\pm 3  \, (22.0\%) \end{array}$ &
$\begin{array}{c}197 \pm 145 \, (>100\%) \end{array}$ &
$\begin{array}{c} 70 \pm 2 \, (3.9\%)\end{array}$ &
$\begin{array}{c} 39 \pm 1 \, (46.7\%)\end{array}$\\ \hline \hline
\end{tabular}
\caption{\footnotesize  $P(D)$ results from the Bayesian/MaxEnt
method for the $2 0 0$ and $4 0 0$ line profiles using different
\emph{a priori} models. The values for $D_{0}$, $\sigma_{0}$,
$\langle D \rangle$ and $\sigma_{\langle D \rangle}^{2}$ were
determined by fitting the Bayesian/MaxEnt solutions with a
log-normal distribution. The percentage difference between
calculated and  theoretical values are given in
parentheses.\label{tab_unip}}
\end{center}
\end{table}

\begin{figure}[thb!]
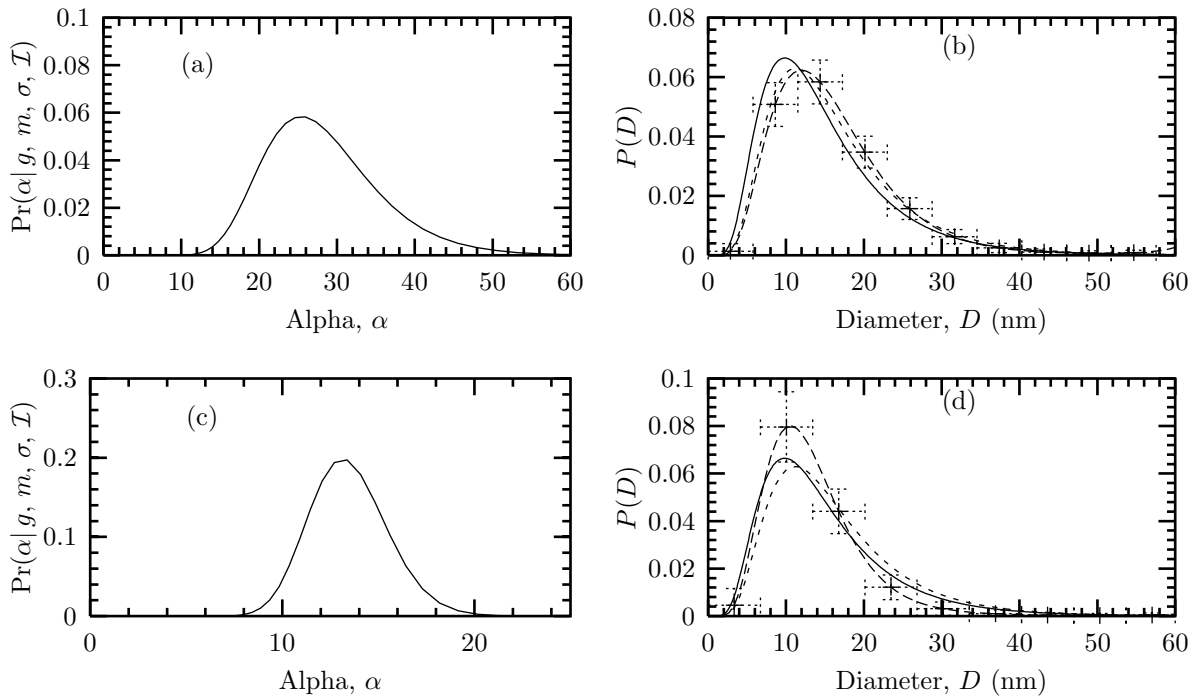

\begin{tabular}{cc}
{\footnotesize\input{size_fig_alpha_sim200non.tex}} &
{\footnotesize\input{size_fig_psol_sim200non.tex}}\\
{\footnotesize\input{size_fig_alpha_sim400non.tex}} &
{\footnotesize\input{size_fig_psol_sim400non.tex}}
\end{tabular}
\caption{\footnotesize   Bayesian/MaxEnt crystallite size
distributions using a log-normal  \emph{a priori} model. (a) The
$\Pr(\alpha|\, \mathbf{g},\,\mathbf{m},\,\sigma,\,\cal{I})$
distribution, (\ref{equ_subequ1}), used to average over the set of
solutions, $\{\mathbf{P}(\alpha)\}$. (b) The theoretical
crystallite size distribution (solid line), Bayesian/MaxEnt size
distribution (long dashed line + error bars) and the low-resolution 
\emph{a priori} model (short dashed line). (c) \& (d) The
corresponding $\Pr(\alpha|\,
\mathbf{g},\,\mathbf{m},\,\sigma,\,\cal{I})$ distribution and
Bayesian/MaxEnt size distribution for the $4 0 0$ line profile.
\label{fig_nonbayessol}}
\end{figure}

\paragraph{`Low resolution' approach}
A log-normal \emph{a priori} model used in the Bayesian/MaxEnt
method was defined from the $D_{0}$ and $\sigma_{0}$ of the $2 0
0$ fuzzy pixel/MaxEnt line profile (see
Table~\ref{tab_memresults}). Unlike the uniform model, this model
defines local features of the size-distribution. The
Bayesian/MaxEnt results using this model are shown in
Fig.~\ref{fig_nonbayessol} and the determined parameters in
Table~\ref{tab_unip}. Before discussing the results, it is
interesting to point out that the log-normal model and theoretical
size-distribution produce a difference of 15.8\%. 
One of the aims of this section is to assess whether this difference has been
imparted to the Bayesian/MaxEnt size-distribution.

Comparing the \emph{a posteriori} distribution for $\alpha$ using
a uniform model (see Figs.~\ref{fig_unibayessol} (a) \& (c)) with
that of the log-normal distribution, given in
Figs.~\ref{fig_nonbayessol}(a) \& (c), we notice that the effect
of the log-normal model is to shift the distribution in
$\alpha$-space and widen it. Essentially the solution space
parameterized by $\alpha$ has been expanded to encompass those
solutions which correspond to the available \emph{a
  priori} and experimental data.

The Bayesian/MaxEnt size distributions, given in
Figs.~\ref{fig_nonbayessol}(b) \& (d), compare reasonably well
with the theoretical distribution. However, there is noticeable
misfitting between these distributions. Further, the
Bayesian/MaxEnt solution has been shifted slightly relative to the
log-normal model. This is also evident in the $R_{f}$ for the $2 0
0$ and $4 0 0$ size distributions, given in Table~\ref{tab_unip}.
The $R_{f}$ for  both solutions has increased relative to the
log-normal model by an additional $\sim 3-6\%$. This can also be
seen by comparing the percentage differences for the $D_{0}$,
$\sigma_{0}$, $\langle D \rangle$ and $ \sigma_{\langle D
\rangle}^{2}$ parameters for the $2 0 0$ fuzzy pixel solution,
given in Table~\ref{tab_memresults} (i.e. third column), with
those given in Table~\ref{tab_unip} using the `low resolution'
method, where there is a slight increase in the percentage
difference, with the exception of the $\sigma_{\langle D
\rangle}^{2}$ value. Additional calculations suggest that
misfitting between the solution and theoretical size
distributions arises from  errors in the \emph{a prior} model. The
influence of the background estimation which was problematic in
the fuzzy pixel analysis does not seem to be a factor in this
analysis.

While there exists some misfitting between the solution and
theoretical size distributions, the vertical error bars correctly
account for this misfitting. This characteristic of the
Bayesian/MaxEnt can be seen for both the uniform and non-uniform
models. Indeed, this feature of the method ensures that it is
fully quantitative, and  represents a clear strength over existing
methods. Comparing these solutions with those using a uniform
model, considerable improvement in the size distribution has been
achieved. The `local' information defined in the log-normal
\emph{a priori} model has been imparted to the Bayesian/MaxEnt
solution.

This analysis also demonstrates the difficulty in
estimating a suitable non-uniform model based on the current
techniques. Further, any uncertainty in the model parameters
is also passed on to the solution distribution. This indicates the
need to quantify the uncertainty in the model parameters and
quantify how these uncertainties are passed on to the solution
size distribution.

\section{Experimental Details}\label{sec_exp}
Analysis of the simulated data highlighted difficulties of
background estimation and the effect of the \emph{a priori} model
on the Bayesian/MaxEnt size distribution. However, this analysis
provided a useful understanding of the experimental condition which
were used in conducting an 
appropriate set of measurements. The fuzzy pixel/Bayesian/MaxEnt methods
were applied to experimental CeO$_2$ diffraction data to determine
the specimen profiles, crystallite shape, and size distribution.
These results are compared with  transmission electron microscopy
data.

\subsection{XRD Details}
The CeO$_2$ specimen used here was prepared for the recent CPD and
IUCR size round robin by \citeasnoun{louer01}.

\begin{figure}[htb!]
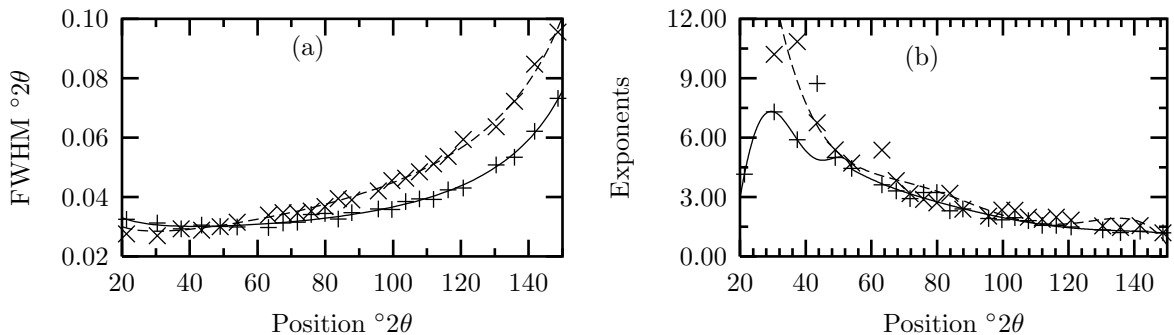

\begin{tabular}{cc}
{\footnotesize\input{size_fig_fwhmplot.tex}} &
{\footnotesize\input{size_fig_expos.tex}}
\end{tabular}
\vspace{-5mm} \caption{\footnotesize  Calibration plots for the
split-Pearson VII line profiles, generated from the SRM 660a
(LaB$_{6}$) diffraction pattern and used to model the instrument
function, $K(2\theta)$. (a) The FWHM vs $2\theta$ for low- ($+$ \&
solid line) and high- ($\times$ \& dashed line) angle sides of the
peak. (b) The exponents vs $2\theta$  for low- ($+$ \& solid line)
and high- ($\times$ \& dashed line) angle sides of the peak.
\label{fig_exp_lab6}}
\end{figure}

Diffraction patterns were collected on a Siemens
D500\footnote{Certain commercial materials, equipment and software
are identified in order to adequately specify the experimental
procedure. Such identification does not imply a recommendation or
endorsement by NIST, nor does it imply that the materials or
equipment or software are necessarily the best available for the
purpose.} diffractometer equipped with a focusing Ge incident beam
monochromator, sample spinner and a scintillation detector. Copper
K$_{\alpha 1}$ radiation with a wavelength $\lambda = 0.15405945
\, \mbox{nm}$ was used. The divergence slit was $0.67\,^{\circ}$,
while the receiving optics included a slit of $0.05\,^{\circ}$ and
$2\,^{\circ}$ Soller slits.  Data were collected in discrete
regions straddling the maxima of each profile, with the step and
scan width of each region being varied in correspondence with the
FWHM.  Count times were varied so as to obtain an approximately
constant total number of counts for each scan region. The
instrument profile function was determined using a split-Pearson
VII profile shape function fitted to 22 reflections collected from SRM
660a (LaB$_{6}$). Fig.~\ref{fig_exp_lab6}, shows the FWHMs and
exponents for the  split-Pearson VII profile function. The low- and
high-FWHMs were fitted using \cite{cheary95}
\begin{equation}
  \mbox{FWHM}^{2}=A \, \tan^{2}\theta + B\,\cot^{2}\theta + C\,\tan
  \theta +D, \label{equ_exp_fwhms}
\end{equation}
while the low- and high-exponents were fitted using a fifth-order
polynomial.

\begin{figure}[htb!]
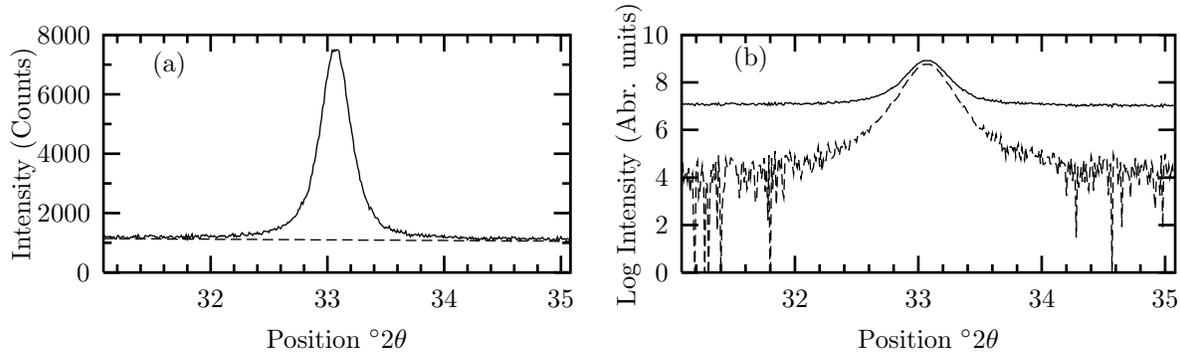

\begin{tabular}{cc}
{\footnotesize\input{size_fig_obs200.tex}} &
{\footnotesize\input{size_fig_logobs200.tex}}
\end{tabular}
\vspace{-5mm} \caption{\footnotesize CeO$_{2}$e Experimental $2 0
0$ profile, $g(2\theta)$: (a) The observed $2 0 0$ profile (solid
line) and estimated background level (dashed line) over
$(2\theta_{0}\pm 2)\,^{\circ}2\theta$, the range over which
the analysis was carried out. (b) Logarithm of the $2 0 0$
measured profile before (solid line) and after the background
estimation (dashed line). \label{fig_exp_ceo2}}
\end{figure}

The count times for the CeO$_2$ data were optimized using
(\ref{equ_errorcounts}) so that the percentage error was kept in
the range 1-3\%  for all peaks in the CeO$_2$ pattern.  The scan
ranges for the CeO$_2$ data were considerably wider, in proportion
to the FWHM, than those used for the data collection from SRM
660a.  This ensured a reasonable determination of the tails of the
profiles and background levels. The CeO$_2$ $2 0 0$ line
profile is shown in
Fig.~\ref{fig_exp_ceo2}(a). This illustrates a typical
experimental line profile using the above conditions and settings.
The estimated (linear) background level is also shown. A log plot
of the $2 0 0$ line before and after the
background estimation is shown in Fig.~\ref{fig_exp_ceo2}(b). The
procedure for determining the background level is as described in
\S~\ref{sec_genmock}.

\subsection{TEM Details}
Particle agglomerates were gently crushed in ethanol using a
mortar and pestle. A portion of the dilute slurry was dispersed on
a holey carbon film and left to dry. Once in the TEM, a series of
micrographs of particles were taken at a fixed magnification of
200k$\times$. In the \emph{preliminary} examination reported here,
these negatives were scanned and analysed by manually
approximating the particle size with an oval. The oval's major and
minor axes were adjusted so as to tangentially intersect the
particle surface facets.

There are several sources of error in the measurements: TEMs typically
have a 5\% error in length scale measurements; also, imaging the
particle clusters means that particles are at different heights, which
results in Fresnel fringes around the particles making it harder to
identify particle edges. Further, larger particles give better
contrast and it is easier to detect their edges, so it is possible to
inadvertently preferentially choose larger particles over smaller
ones.

A frequency histogram for about 850 particles is shown in
Fig.~\ref{fig_pnon}(d). This figure also shows the Bayesian/MaxEnt
size distributions (discussed in \S~\ref{sec_expquant}) determined
from the non-overlapped $h k l$ profiles of the CeO$_2$
diffraction pattern. A TEM micrograph of the CeO$_2$ crystallites
is shown in Fig.~\ref{tem_images}. It can be seen from the larger
particles that they have a spherical-like morphology.

\section{Analysis of  CeO$_2$ x-ray diffraction data}

Two levels of application of the Bayesian and MaxEnt theory has
been chosen in our analysis. We refer to these as the
\emph{qualitative} and \emph{quantitative} approaches, to reflect
their degree of rigor (see \S~\ref{sec_expqual} \&
\ref{sec_expquant}, respectively).

\subsection{Qualitative analysis}\label{sec_expqual}
The qualitative analysis is used to determine the type and nature
of specimen broadening, by first determining the specimen profile,
$f$, followed by the application of the Warren-Averbach and
Williamson-Hall methods. The integral breadths, from a
Williamson-Hall plot, identify the presence of both strain- and
size-broadening contributions, while plotting multiple-order
Fourier coefficients and all other available Fourier coefficients
on the same axes, also allows size- and strain-broadening
contributions be identified \citeaffixed{armstrong99}{see}.

We have introduced the fuzzy pixel/MaxEnt method for determining
$f$ to ensure that no artifacts (such as spurious oscillations in
the tails of $f$) are promulgated to the solution, and also to
preserve the positivity of $f$.

We stress that unlike traditional methods, the approach in this
section makes no assumptions at all about the nature of the
specimen profile or broadening (i.e be it Gaussian, Lorentzian,
Voigtian etc.). Thus, in further distinction from  traditional
deconvolution approaches, our approach facilitates the subsequent
unbiased assessment of anisotropic broadening in the specimen, for
example using contrast factors \cite{ungar99a}.

\begin{figure}[htb!]
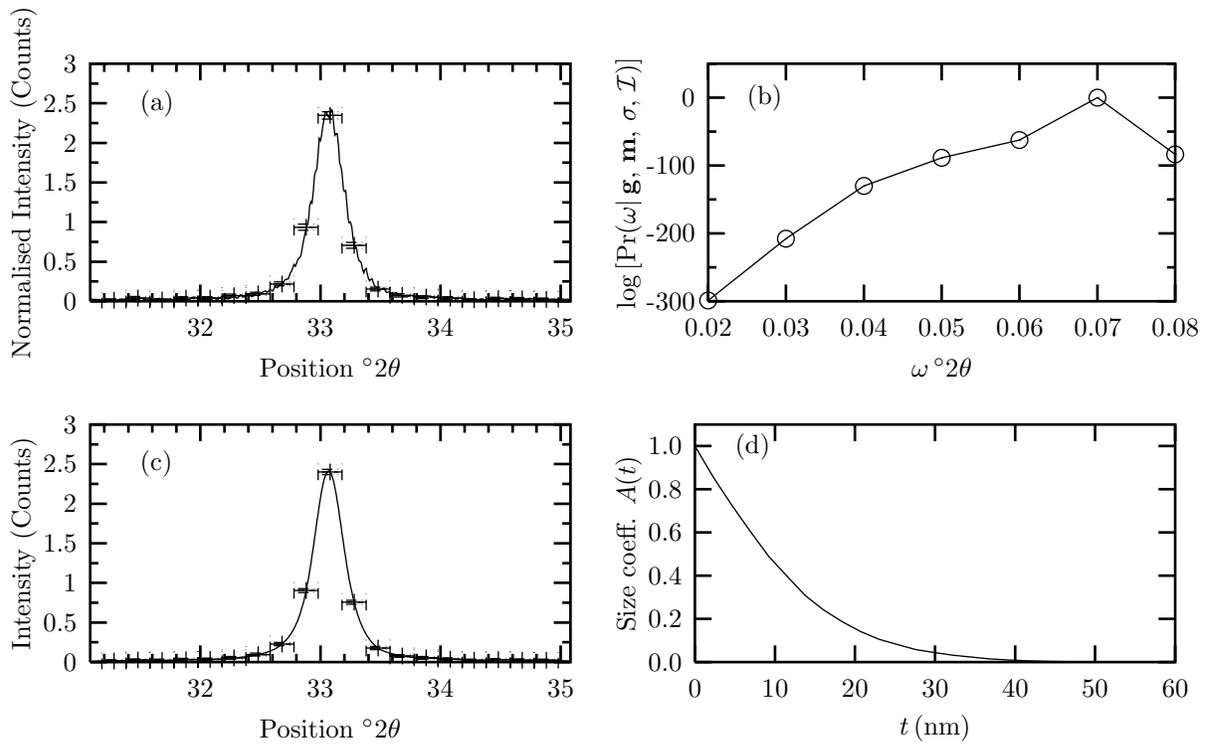

\begin{tabular}{cc}
\footnotesize{\input{size_fig_oldfsol_exp200.tex}}
&\footnotesize{\input{size_fig_fuzzyw_exp200.tex}}\\
\footnotesize{\input{size_fig_fuzzyfsol_exp200.tex}} &
\footnotesize{\input{size_fig_fuzzycoeffs_exp200.tex}}
\end{tabular}
\caption{\footnotesize Specimen profiles from the `old' MaxEnt
method and fuzzy pixel/MaxEnt method for the measured CeO$_2$
$2 0 0$ line. (a) `Old MaxEnt' specimen profile (solid line +
error bars). (b) $\log
\Pr(\omega|\,\mathbf{g},\,\mathbf{m},\,\sigma,\,\cal{I})$
distribution to determine the optimum fuzzy pixel width,
$\hat{\omega} \approx 0.07\,^{\circ}2\theta$. (c) Fuzzy
Pixel/MaxEnt specimen profile (solid line + error bars). (d) Fuzzy
Pixel/MaxEnt Fourier coefficients (solid line).
\label{fig_expfuzzsol}}
\end{figure}

Fig.~\ref{fig_expfuzzsol} shows an example of the fuzzy
pixel/MaxEnt method applied to the CeO$_2$ measured $2 0 0$
line profile given in Fig.~\ref{fig_exp_ceo2}.
Fig.~\ref{fig_expfuzzsol}(a) is an example of the `old' MaxEnt
method, showing the effect of noise amplification. On applying the
fuzzy pixel/MaxEnt method, the correlation length scale for the
profile was determined, as discussed in \S~\ref{sec_fuzzy} and is
shown in Fig.~\ref{fig_expfuzzsol}(b); the subsequent $f$ and
Fourier coefficients for the $2 0 0$ line profile are given in
Figs.~\ref{fig_expfuzzsol}(c) \& (d), respectively. As
demonstrated in the analysis of the simulated data, there is
noticeable improvement in the quality of the solution line profile
using the fuzzy pixel/MaxEnt method. This approach was applied to
all the non-overlapped line profiles, including $1 1 1$, $2 0 0$,
$2 2 0$, $4 0 0$, $4 2 2$, $5 1 1$ and $5 3 1$.

\begin{figure}[htb!]
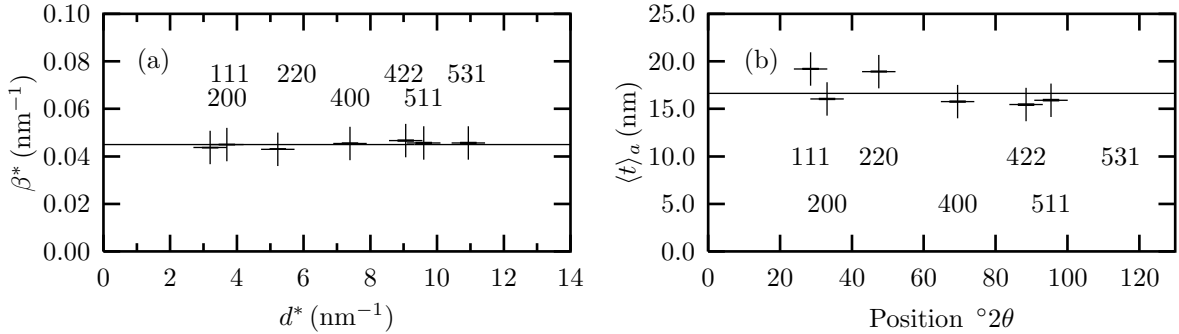

\begin{tabular}{cc}
{\footnotesize\input{size_fig_whplot.tex}} &
{\footnotesize\input{size_fig_areasizes.tex}}
\end{tabular}
\vspace{-5mm} \caption{\footnotesize Volume- and area-weighted
sizes from the Fuzzy Pixel/MaxEnt line profiles for measured
CeO$_2$ data. (a) Classical Williamson-Hall plot of the integral
breadths, showing no dependency on $h k l$. This suggests that the
crystallites are spherical in shape.  (b) Area-weighted sizes
determined from the Fourier coefficients of the Fuzzy Pixel/MaxEnt
line profiles. \label{fig_exp_sizes}}
\end{figure}

The volume- and area-weighted sizes were determined from the
Williamson-Hall plot and Fourier coefficients, respectively. These
results are shown in Fig.~\ref{fig_exp_sizes} and summarized in
Table~\ref{tab_expsizes}.

\begin{table}[htb!]
\tiny
\begin{center}
\begin{tabular}{ccccccccc}\hline \hline
$h k l$  & $\begin{array}{c} \langle t \rangle_{a} \\ \mbox{(nm)}
  \end{array}$ &  $\begin{array}{c}\langle D \rangle_{a} \\
    \mbox{(nm)} \end{array}$ &  $\begin{array}{c} \langle t
  \rangle_{v} \\  \mbox{(nm)} \end{array}$ &  $\begin{array}{c}\langle
  D \rangle_{v} \\ \mbox{(nm)} \end{array}$ & $\begin{array}{c} D_{0}
  \\ \mbox{(nm)} \end{array}$ & $\sigma_{0}$ &
  $\begin{array}{c} \langle D \rangle \\  \mbox{(nm)}\end{array}$ &
  $\begin{array}{c} \sigma_{\langle D \rangle}^{2} \\ (\mbox{nm}^{2})
  \end{array}$ \\ \hline
$1 1 1$ & $19.21 \pm 0.05$ & $28.81 \pm 0.07$ & $22.88 \pm 0.03$ &
$30.50 \pm 0.04$ & $25.0 \pm 0.2$ & $1.270 \pm 0.007$ & $25.7\pm  0.2$
& $39 \pm  2$ \\
$2 0 0$ & $16.04 \pm 0.06$ & $24.06 \pm 0.08$ & $22.22 \pm 0.05$ &
$29.63 \pm 0.06$ & $14.3 \pm 0.2$ & $1.578 \pm 0.007$ & $15.9 \pm 0.2$ &
$58 \pm 2$ \\
$2 2 0$ & $18.92 \pm 0.04$ & $28.38 \pm 0.06$ & $23.24 \pm 0.04$ &
$31.00 \pm 0.05$ & $22.8 \pm 0.2$ & $1.345 \pm 0.006$ & $23.8 \pm 0.2$ &
$52 \pm 2$ \\
$4 0 0$ & $15.76 \pm 0.06$ & $23.64 \pm  0.09$ & $22.03 \pm 0.11$ &
$29.4 \pm  0.2$ & $13.7 \pm  0.3$ & $1.59 \pm 0.01$ & $15.3 \pm 0.3$ &
$56 \pm 3$\\
$4 2 2$ & $15.45 \pm 0.08$ & $23.2 \pm 0.1$ & $21.5 \pm 0.1$ &
$28.6 \pm 0.1$ & $13.7 \pm 0.3$ & $1.58 \pm 0.01$ & $15.2 \pm 0.3$
& $54 \pm 3$\\
$5 1 1$ & $15.91 \pm  0.07$ & $23.9  \pm  0.1$ & $21.9 \pm 0.1$ &
$29.2 \pm  0.2$ & $14.4 \pm  0.3$ & $1.57 \pm 0.01$ & $16.0 \pm
0.3$ & $57 \pm  3$ \\
$5 3 1$ & $15.04 \pm  0.04$ & $22.55 \pm  0.07$ & $21.9 \pm 0.1$ &
$29.20 \pm 0.16$ & $11.8 \pm  0.2$ & $1.66 \pm  0.01$ &
$13.5 \pm 0.2$ & $53 \pm 3$\\ \hline
Average & $16.6 \pm 0.2$ & $24.9 \pm 0.2$ & $22.2 \pm 0.2$ & $29.6 \pm 0.3$ &
$16.5 \pm  0.6$ & $1.51 \pm 0.03$ & $17.9 \pm  0.7$ & $53 \pm  7$ \\ \hline \hline
\end{tabular}
\caption{\footnotesize Summary of CeO$_2$ data analysis. The area-
and volume-weighted sizes were determined from the specimen
profile of the fuzzy pixel/MaxEnt method. The $\langle t
\rangle_{a}$ and $\langle t \rangle_{v}$ results were determined
directly from $f$ using (\ref{equ_areat}) and
(\ref{equ_intbreadth}), respectively. The area- and
volume-weighted diameters were determined using (\ref{equ_diam}),
while the log-normal parameters were determined from Krill \&
Birringer (1998) and using
(\ref{equ_lognormal}).\label{tab_expsizes}}
\end{center}
\end{table}

Fig.~\ref{fig_exp_sizes}(a) shows the Williamson-Hall plot for the
non-overlapped line profiles. It is evident that  size effects
are the dominant source of specimen broadening, since there is no
detectable slope in the integral breadth data. Moreover, there is
no systematic variation of the integral breadths with $h k l$,
further suggesting that the crystallite shape is independent of $h
k l$. From these results, we can infer that the average shape of
the crystallites is spherical. This is further supported by the
area-weighted sizes shown in Fig.~\ref{fig_exp_sizes}(b). These
results were determined by applying (\ref{equ_areat}) to the
Fourier coefficients of the fuzzy pixel/MaxEnt specimen profiles
and plotted over the entire $2\theta$-range. Again, the relative
uniformity of this plot suggests that size effects are the major
source of specimen broadening and that crystallites are
near-spherical in shape. Deviations for the $1 1 1$ and $2 2 0$
data points in Fig.~\ref{fig_exp_sizes}(b) arise from the
differentiation of (\ref{equ_areat}) in the region $t \rightarrow
0$, where perturbations in the Fourier coefficients cause large
changes in the area-weighted size \cite{armstrong99}. In addition,
the Fourier coefficients for all the non-overlapped $h k l$ lines
suggest that the maximum crystallite size is
$\sim50-60\,\mbox{nm}$. An example of this can be seen in
Fig.~\ref{fig_expfuzzsol}(d), where $A(t) \sim 0$ for
$\sim50-60\,\mbox{nm}$. This can also be seen from the discussion
in \S~\ref{sec_sizebroadening} and by inspecting
Fig.~\ref{fig_common}, where the boundary conditions for $A(t)$
(or  $V(t)$) are defined in terms of the maximum size in the
direction of the scattering vector.

Referring to Table~\ref{tab_expsizes}, a spherical crystallite
shape model was used to determine the  area- and volume-weighted
diameters, together with (\ref{equ_areadiam}) and
(\ref{equ_voldiam}), respectively. The log-normal distribution
parameters, $D_{0}$, $\sigma_{0}$, $\langle D \rangle$ and
$\sigma_{\langle D \rangle}^{2}$ were determined using the
equations developed by \citeasnoun{krill98} and
(\ref{equ_lognormal}), which relate the log-normal parameters to
the area- and volume-weighted sizes and the average diameter,
$\langle D\rangle$, and variance $\sigma_{\langle D \rangle}$.

It can be seen from Fig.~\ref{fig_expfuzzsol} and
Table~\ref{tab_expsizes}, that the area- and volume-weighted sizes
are relatively uniform for the $2\theta$ (or $h k l$) range. The
quoted uncertainty for the averages was determined from a sum of
squares of the uncertainties in the tabulated results.

The average results for $D_{0}$ and $\sigma_{0}$, were used to
define a log-normal \emph{a priori} model in the Bayesian/MaxEnt
method (see \S~\ref{sec_expquant}). By defining the  \emph{a
priori} model as a log-normal distribution, we are essentially
testing the assumption that the size distribution is a log-normal
distribution.

If the underlying size distribution is indeed log-normal, with
parameters close to those in Table~\ref{tab_expsizes}, then we
would expect the  Bayesian/MaxEnt solution  to lie `close' to the
\emph{a priori} model. However, if the Bayesian/MaxEnt solution
were `some distance' from  the \emph{a priori} model, this would
imply that either the underlying parameters or the model were
inappropriately defined.  The former case was demonstrated in
analysis of the simulated data (see \S~\ref{sec_mockanal}), where
uncertainties in the log-normal model were passed onto the
Bayesian/MaxEnt solution; the latter case requires additional
Bayesian analysis to test possible models \cite{sivia93,loredo01}.

In summary, the qualitative analysis has applied the fuzzy
pixel/MaxEnt method to determine the specimen profile $f$ for
all non-overlapped line profiles from the CeO$_{2}$ measured
data (see Fig.~\ref{fig_expfuzzsol}). This enabled subsequent
analyses to determine the Fourier coefficients, integral breadths,
and the area- and volume-weighted sizes. Fig.~\ref{fig_exp_sizes}
and Table~\ref{tab_expsizes} clearly indicate that the CeO$_{2}$
specimen on average consists of spherical crystallites. While a log-normal
distribution can be fitted to these results, a quantitative method
such as the Bayesian/MaxEnt technique is needed to determine the
CeO$_{2}$ size distribution \emph{directly} from the experimental
data and to verify the assumption of a log-normal model.

\subsection{Quantitative analysis}\label{sec_expquant}
The quantitative analysis method uses the \emph{a priori}
information determined from the qualitative analysis and the
available experimental data (such as the instrument and profile
kernels, statistical uncertainties and experimental line profiles)
to directly determine the crystallite size distribution.

The MaxEnt method also enables an \emph{a priori} model to be
included, while quantifying the uncertainty in the solution size
distribution.

In this section, we apply the Bayesian/MaxEnt method to the
CeO$_{2}$ data. The analysis presented here follows
the steps discussed in \S~\ref{sec_mockanal}. Two \emph{a
  priori} models are used: -- (i) a uniform model, and (ii) the
log-normal distribution determined in \S~\ref{sec_expqual}. The
Bayesian/MaxEnt size distributions for each case are fitted with a
log-normal distribution, while the size distributions from (ii)
are compared with the TEM size distribution, with very good
agreement.

\paragraph{Uniform model}
A uniform model was defined over the region of $D\in
[0,\,60]\,\mbox{nm}$ determined by the Fourier coefficients of the
specimen profile, where $A(t) \sim 0$.  This is illustrated by the
Fourier coefficients for the $2 0 0$ line profile, given in
Fig.~\ref{fig_expfuzzsol}(d). The Bayesian/MaxEnt size
distributions using this model are shown in Fig.~\ref{fig_puni}
for the $2 0 0$ line profile (see Fig.~\ref{fig_puni} (a \& b)).
The size distributions for the non-overlapped line profiles are
given in Fig.~\ref{fig_puni} (c \& d).

\begin{figure}[thb!]
\begin{tabular}{cc}
\footnotesize{\input{size_fig_alpha_exp200uni.tex}} &
\footnotesize{\input{size_fig_pdist_exp200uni.tex}}\\
\footnotesize{\input{size_fig_pdist_uni.tex}} &
\footnotesize{\input{size_fig_avedia_exp.tex}}
\end{tabular}
\caption{\footnotesize CeO$_2$ Bayesian/MaxEnt crystallite-size
distribution using a uniform \emph{a priori} model over
$D\in[0,\,60]\,\mbox{nm}$: (a) The
$\Pr(\alpha|\,\mathbf{g},\,\mathbf{m},\,\sigma,\,\cal{I})$
distribution, (\ref{equ_subequ1}), used to average over the set of
solutions, $\{\mathbf{P}(\alpha)\}$  for the $2 0 0$ line profile;
(b) Bayesian/MaxEnt crystallite-size distribution (solid line +
error bars) and the \emph{a priori} model over
$D\in[0,\,60]\,\mbox{nm}$; (c) CeO$_2$ Bayesian/MaxEnt
crystallite-size distributions for the various CeO$_2$ $h k l$
profiles; (d) Average diameters from the uniform model ($+$) and
log-normal models ($\times$). The horizontal lines represent the
average for each model.\label{fig_puni}}
\end{figure}
\begin{table}[htb!]
\footnotesize
\begin{center}
\begin{tabular}{ccccc}\hline\hline
 $h k l$   & $\begin{array}{c} D_{0} \\ \mbox{(nm)} \end{array}$ & $\sigma_{0}$ &
  $\begin{array}{c} \langle D \rangle \\  \mbox{(nm)}\end{array}$ &
  $\begin{array}{c} \sigma_{\langle D \rangle}^{2} \\ (\mbox{nm}^{2})
  \end{array}$ \\ \hline
$1 1 1$ & $13.1 \pm 0.4$ & $1.9 \pm 0.1$ & $16.3 \pm 0.8$ & $140 \pm 43$ \\
$2 0 0$ & $14.7 \pm 0.3$ & $1.61 \pm 0.03$ & $16.4 \pm 0.4$ & $69 \pm 7$ \\
$2 2 0$ & $ 15.5 \pm 0.4$ & $1.70 \pm 0.08$ & $17.8 \pm 0.6$ & $104\pm 25$ \\
$4 0 0 $ & $15.8 \pm 0.8$ & $1.7 \pm 0.1$ & $18 \pm 1$ & $120 \pm 48$ \\
$4 2 2 $ & $11.1 \pm 0.2$ & $1.728 \pm 0.007$ & $12.8 \pm 0.2$ & $58\pm 2$ \\
$5 1 1$ & $14.3 \pm 0.3$ & $1.66 \pm 0.05$ & $16.3 \pm 0.4$ & $78 \pm
15$\\ \hline
Average & $14 \pm 1$ & $1.7\pm 0.2$   &$16 \pm 2$ & $95 \pm 71$ \\
\hline \hline
\end{tabular}
\caption{\footnotesize Size distribution results using a uniform
\emph{a priori} model in the Bayesian/MaxEnt method. The
Bayesian/MaxEnt size distributions given in Fig.~\ref{fig_puni}(c)
were fitted with a log-normal size distribution and the above
parameters determined.\label{tab_expsizes2}}
\end{center}
\end{table}

The uncertainties in the Bayesian/MaxEnt size distribution for the
$2 0 0$ line profile indicate how little useful \emph{a priori}
information has been transferred from the uniform model to the
final distribution. We also notice that the final distribution is
some distance from the model, illustrating that the underlying
CeO$_2$ crystallite size distribution consists of a non-uniform
structure. As can be seen in Fig.~\ref{fig_puni}(c), the size
distributions are poorly defined in the range of
$D\in[0,\,5]\,\mbox{nm}$; while for $D\gtrsim 5\,\mbox{nm}$ the
non-uniform structure is evident. Since the size distribution is
the only invariant quantity, we also expect the solution for each
$ h k l$ to be the same. From the size distributions given in
Fig.~\ref{fig_puni}(c), there is a broad agreement between the
distributions, with the exception of the $1 1 1$ and $4 2 2$ cases.
Both of these distributions lie at the extremities of the
diffraction pattern and are more likely to be susceptible to
larger experimental uncertainties.

The Bayesian/MaxEnt size distributions were fitted with a
log-normal model and the $D_{0}$, $\sigma_{0}$, $\langle D
\rangle$ and $\sigma_{\langle D \rangle}^{2}$ parameters
determined. These results are given in Table~\ref{tab_expsizes2}.
The uncertainties in the solution distributions for the uniform
model are also reflected in the uncertainties in the fitted
quantities. This is especially the case for the variance of the
size distributions, $\sigma_{\langle D
  \rangle}^{2}$, with an error of $\sim 80\%$. This large uncertainty is a
consequence of the scatter of size distributions shown in
Fig.~\ref{fig_puni}(c). Such scatter is also noticeable when the
average diameters, $\langle D \rangle$,
(Table~\ref{tab_expsizes2}) are plotted, as shown in
Fig.~\ref{fig_puni}(d). The average values for $D_{0}$,
$\sigma_{0}$, $\langle D\rangle$ and $\sigma_{\langle D
\rangle}^{2}$ are again in broad agreement with results determined
in \S~\ref{sec_expqual}, once the uncertainties are taken into
account.

In summary, the use of the uniform model in the Bayesian/MaxEnt
method has shown that there is a non-uniform structure to the
CeO$_2$ size distributions. However, the lack of information in this
model results in large uncertainties and considerable scatter of
the distributions when plotted on the same axes (see
Fig.~\ref{fig_puni}(c)).

\paragraph{Log-normal model}
The parameters for the log-normal distribution determined in
\S~\ref{sec_expqual} were used as the non-uniform \emph{a priori}
model in the Bayesian/MaxEnt method. The model was defined over
the range of $D\in [0,\,60]\, \mbox{nm}$.

\begin{figure}[thb!]
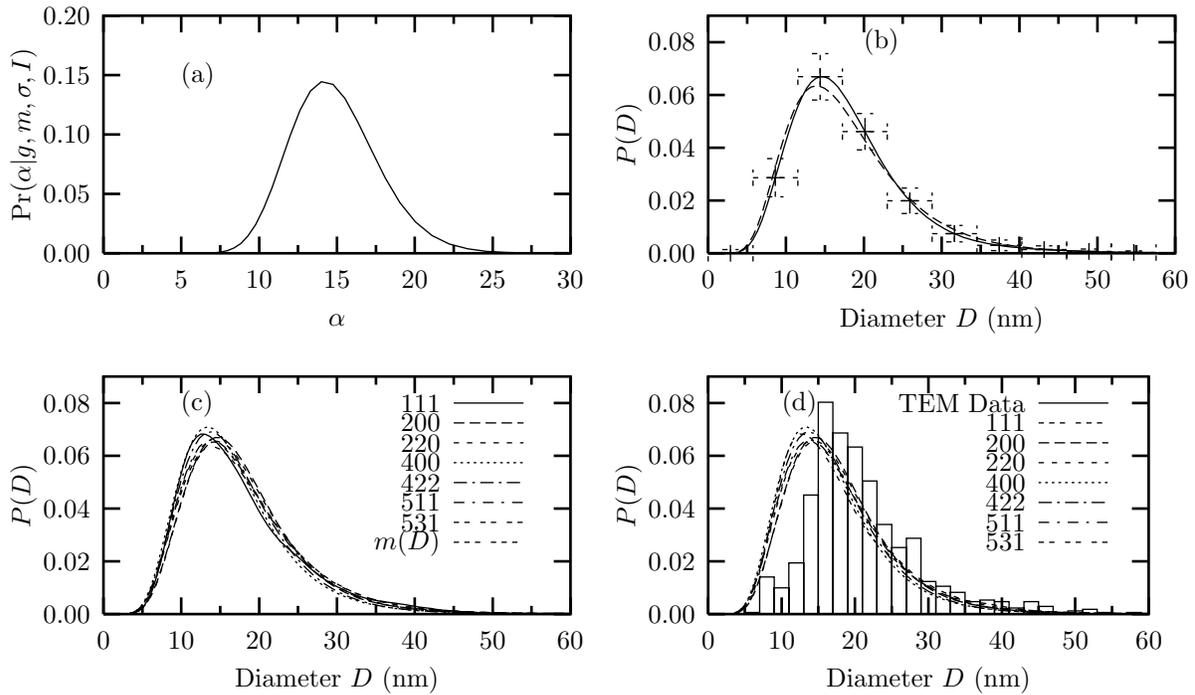

\begin{tabular}{cc}
\footnotesize{\input{size_fig_alpha_exp200non.tex}} &
\footnotesize{\input{size_fig_pdist_exp200non.tex}}\\
\footnotesize{\input{size_fig_pdist_non.tex}} &
\footnotesize{\input{size_fig_tem.tex}}
\end{tabular}
\caption{\footnotesize CeO$_2$ Bayesian/MaxEnt crystallite-size
distribution using a log-normal \emph{a priori} model over
$D\in[0,\,60]\,\mbox{nm}$. (a) $\Pr(\alpha|\,
\mathbf{g},\,\mathbf{m},\,\sigma,\,\cal{I})$ distribution,
(\ref{equ_subequ1}), used to average over the set of solutions
$\{\mathbf{P}(\alpha)\}$,  for the $2 0 0$ line. (b)
Bayesian/MaxEnt crystallite-size distribution (solid line + error
bars) and the \emph{a priori} model (dashed line) over
$D\in[0,\,60]\,\mbox{nm}$. (c) CeO$_2$ Bayesian/MaxEnt
crystallite-size distributions for the non-overlapped line
profiles. (d) Comparison of TEM and Bayesian/MaxEnt size
distributions. Due to difficulties in de-aggregating the CeO$_{2}$
particles and identifying smaller crystallites, the TEM size
distribution is poorly defined for $D\leq 14\,\mbox{nm}$. However,
for $D\geq 14\,\mbox{nm}$, the comparison is
excellent.\label{fig_pnon}}
\end{figure}

The Bayesian/MaxEnt size distributions for this model are shown in
Fig.~\ref{fig_pnon}. The results are listed in
Table~\ref{tab_expsizes3}. Figs.~\ref{fig_pnon}(a) \& (b) show the
results for the $2 0 0$ size distribution using this model. The
Bayesian/MaxEnt solution lies close to the log-normal model,
while the uncertainties have decreased considerably compared with
the size distribution (using a uniform model) in
Fig.~\ref{fig_puni}(b); however, although the vertical error bars
have decreased, they are still considerable. This can be explained
in terms of the influence of the peak-to-background ratio. As
discussed in \S~\ref{sec_mockfuzzy}, the variance of the
experimental data is determined by two terms, the statistical
noise and the variance on the estimated background level. If the
peak-to-background ratio is large ($\gtrsim 10$), then the
statistical noise dominates and the corresponding error bars in
the Bayesian/MaxEnt distribution become small when the solution is
close to the underlying size distribution. This has been
demonstrated using computer simulations. However, if the
peak-to-background ratio is finite ($< 10$), the corresponding
error bars in the MaxEnt/Bayesian solution  remain finite
regardless of how close the solution is to the underlying
distribution. This is a direct consequence of determining the size
distribution directly from the experimental data.

The Bayesian/MaxEnt size distributions for all the non-overlapped
$h k l$ line profiles are shown in Fig.~\ref{fig_pnon}(c). They
lie very close to each other, reflecting the invariance of the
size distribution and remaining close to the  log-normal model.
The scatter in the size distributions that was noticeable in
Fig.~\ref{fig_puni}(c) for the uniform model has disappeared.
Further, these results imply that the underlying size distribution
from the CeO$_2$ crystallites can be described by a log-normal
distribution. Comparing these results with the TEM size
distribution, very good agreement is obtained for $14 \lesssim
D\lesssim 60 \,\mbox{nm}$. Due to its poor statistics, the TEM
size distribution is ill-defined for $D\lesssim 14\,\mbox{nm}$.
As mentioned above, the CeO$_{2}$ agglomerates were not separated,
making it difficult to identify the smaller crystallites and
contributing to the poorly defined region for $D\lesssim
14\,\mbox{nm}$.  The TEM size distribution given in
Fig.~\ref{fig_pnon}(d), represents a \emph{preliminary} set of
data and further results are currently being collated.

\begin{table}[htb!]
\footnotesize
\begin{center}
\begin{tabular}{ccccc}\hline \hline
 $h k l$   & $\begin{array}{c} D_{0} \\ \mbox{(nm)} \end{array}$ & $\sigma_{0}$ &
  $\begin{array}{c} \langle D \rangle \\  \mbox{(nm)}\end{array}$ &
  $\begin{array}{c} \sigma_{\langle D \rangle}^{2} \\ \mbox{(nm$^2$)}
  \end{array}$ \\ \hline
$1 1 1$ & $15.9 \pm  0.2$ & $1.505\pm 0.008$ & $17.2\pm  0.2$ &$54 \pm 2$\\
$2 0 0$ & $16.64 \pm 0.04$ & $1.469 \pm 0.005$ & $17.91 \pm 0.04$ &$51 \pm 1$\\
$2 2 0$ & $15.68 \pm 0.06$ & $1.502 \pm  0.008$ & $17.04\pm 0.08$ & $52 \pm 2$\\
$ 4 0 0$ & $15.48 \pm  0.01$ & $1.480 \pm  0.005$ & $16.72 \pm
0.03$ & $46 \pm 1$\\
$4 2 2$ &  $15.86 \pm  0.07$ & $1.4799 \pm 0.0002$ & $17.12 \pm
0.08$ & $48.7 \pm   0.5$ \\
$5 1 1$ & $16.21 \pm  0.07$ & $1.497 \pm 0.002$ & $17.58\pm 0.08$ & $54.7 \pm 0.6$\\
$5 3 1$ & $16.32 \pm 0.07$ & $1.500 \pm 0.005$ & $17.72 \pm 0.09$ &
$56 \pm 1$ \\ \hline
Average & $16.0 \pm  0.2$ & $1.49 \pm 0.01$ & $17.3 \pm 0.3$ & $52
\pm 3$ \\
\hline \hline
\end{tabular}
\caption{\footnotesize Size distribution results using a log-normal
  \emph{a priori}
model in the Bayesian/MaxEnt method. The  Bayesian/MaxEnt size
distributions given in Fig.~\ref{fig_puni}(c) were fitted with a
log-normal size distribution and the above parameters
determined.\label{tab_expsizes3}}
\end{center}
\end{table}
The correspondence between the Bayesian/MaxEnt size distributions
and the TEM distribution is very good for $D \geq 14\,\mbox{nm}$.
The size distributions shown in Fig.~\ref{fig_pnon}(c) were fitted
with a log-normal distribution and the $D_{0}$, $\sigma_{0}$,
$\langle D \rangle$ and $\sigma_{\langle D \rangle}^{2}$
parameters were determined. These results are shown in
Table~\ref{tab_expsizes3}. The fitted distribution compared very
closely with the solution distribution. The small uncertainties in
the fitted quantities of Table~\ref{tab_expsizes3} reflect the
quality of the Bayesian/MaxEnt distributions. This can also be seen
in the low uncertainty in the variance, $\sigma_{\langle D
\rangle}^{2}$, which is $\sim 8 \%$.

The average quantities given in Table~\ref{tab_expsizes3} can be
considered to represent the size distribution for the CeO$_{2}$
specimen. Hence, the use of the fuzzy pixel/Bayesian/MaxEnt
methods has determined the specimen profile, $f$, and enabled size
effects to be identified as the major source of specimen
broadening. The analysis of the line profiles has shown that the
crystallite shape is spherical, on average. The Fourier coefficients of
the specimen profiles also show that the crystallites have a
maximum size of $\sim 60 \,\mbox{nm}$. This was subsequently shown
from the Bayesian/MaxEnt size distributions. Using this
information, the Bayesian/MaxEnt method successfully determined
the CeO$_{2}$ size distribution. While the size distributions
using a uniform \emph{a priori} model broadly agree with the
results from the fuzzy pixel analysis, the uncertainty in the
results is large; on using a log-normal \emph{a priori}
model considerable improvements in the size distribution were
obtained. The non-uniform structure in the model has been transferred to the Bayesian/MaxEnt solution.

\begin{figure}
\begin{center}
\vspace{-70mm}\includegraphics[scale=1.0]{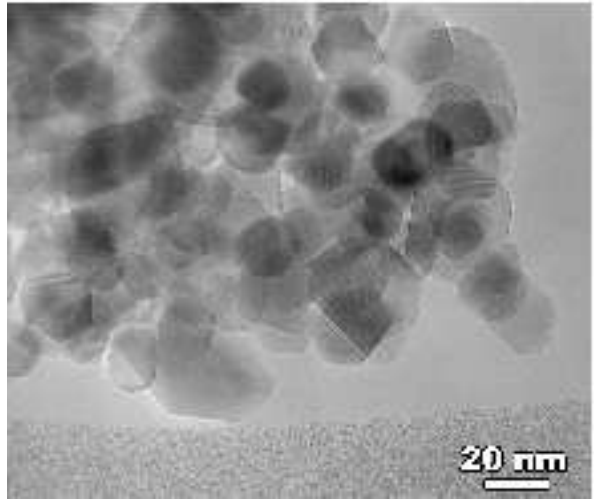}
\caption{\footnotesize A TEM micrograph of the CeO$_{2}$ specimen taken at a
  magnification of $200\,\mbox{k}\times$. The crystallites appear to
  have a spherical-like shape and size that are in the range predicted
  by the crystallite-size analysis presented here. }
\label{tem_images}
\end{center}
\end{figure}

The TEM micrograph of the CeO$_{2}$ specimen, shown in
Fig.~\ref{tem_images}, confirms the results that have been
determined from the x-ray diffraction data. From the micrograph,
it can be seen that the crystallites are near-spherical in shape.
It can also be seen that the crystallites are in the range of size
predicted by crystallite-size analysis. Considerable overlapping of the crystallites, which
complicates the task of gathering sufficiently reliable data for
the TEM size distribution is evident.

\section{Conclusion}
The central aim of this study was to develop a single and
self-contained method for determining the crystallite-size
distribution and shape from experimental line profile data. We
have shown that the crystallite-size distribution can be
determined without assuming a functional form for the size
distribution, determining instead the size distribution with the
least assumptions.

This was achieved by reviewing size broadening theory showing how
the observed line profile can be expressed in terms of the
instrument kernel, line profile kernel and size distribution. It
was also shown that the instrument and line profile kernels could
be combined into a single kernel, hence enabling the simultaneous
removal of  instrumental broadening while determining the size
distribution (see \S~\ref{theory_broadening}).

The development of this method made use of two fundamental
observations--- that distributions such as the specimen profile and
size distribution must be \emph{both} positive and additive. Drawing on
extensive theoretical developments, the entropy function was
selected as the function that can attribute values to the specimen
line profile and size distribution, while preserving the
positivity and additivity of the profile and distribution. It can
be also argued that the entropy function is the only function that
produces consistent results in the light of experimental data (see
\S~\ref{sec_observations}).

Using the mathematical and statistical foundations of Bayesian
theory, the \emph{a posteriori} distributions of $P(\mathbf{D})$
in terms of the experimental data, statistical noise and
scattering kernel can be determined. By maximizing this
distribution, the most probable size distribution can be
calculated from the experimental line profile, without making any
assumptions concerning the functional form of the size
distribution. Determining the most probable size distribution
addresses the inherent non-uniqueness and ill-conditioning in the
integral equations arising from scattering and instrumental
broadening. The generality of this formalism enables any
crystallite shape to be used and any number of principal axes,
$\mathbf{D}=\{D_{1},\,D_{2},\,D_{3}\}$, of the crystallite shape
can be included in determining the corresponding size
distributions.

Simulated data were used to test the fuzzy pixel and
Bayesian/MaxEnt methods on size-broadened line profiles. The
reliability of these methods was established by showing that they
can reproduce the underlying parameters of the area- and
volume-weighted sizes, and the parameters of the size
distributions.

The application of these methods to  CeO${_2}$ experimental data
generally produced very good results. The line profile analysis
applying fuzzy pixel/MaxEnt methods produced reliable and
consistent results over a wide range of low-, mid- and high-angle
profiles.

The application of the Bayesian/MaxEnt method to the CeO$_{2}$
data demonstrated that this method can determine size
distributions, while making the minimum number of assumptions. The
use of a uniform \emph{a priori} model produced broadly consistent
results with the fuzzy pixel/MaxEnt method; however, the lack of
information defined in this model was evident in the large
uncertainties of the estimated quantities.

Using the fuzzy pixel/MaxEnt results as the log-normal \emph{a
priori} model demonstrated that once `useful' information is
encoded in the model, improvements in the size distributions and
considerable reduction in the uncertainties can be achieved.
Analysis of the x-ray diffraction profiles using the log-normal
model in the Bayesian/MaxEnt method revealed that the crystallites
are spherical in shape, with a size distribution corresponding to
the distribution in Fig.~\ref{fig_pnon} and average quantities in
Table~\ref{tab_expsizes3}. The comparison of these Bayesian/MaxEnt
results with TEM results is favorable, but it does reveal
shortcomings in the collected TEM data arising from particle
aggregation. The TEM distribution micrographs support the results
from the line profile analysis.

The use of simulated and experimental data demonstrates that the
fuzzy pixel/Bayesian/ MaxEnt methods are fully quantitative in
their ability to determine and attribute errors to the solution
line profiles and size distributions.

Although the results from the Bayesian/MaxEnt method are in good
agreement and address the limitations of the earlier work
\citeaffixed{armstrong99b,armstrong99}{see}, several important
issues have been raised and are the subject of further
investigation. These concern the accurate background estimation of
the observed line profile and are very important; for example, the
analysis of simulated data demonstrated how systematic errors
affect the Fourier coefficients. Recently, \citeasnoun{david01}
have developed a Bayesian technique for estimating the background,
which can be adopted in this method. Another problem encountered
was in the estimation and quantifying of a non-uniform \emph{a
priori} model. In this analysis we have used the information
determined from the fuzzy pixel/MaxEnt method; however, the issue
of determining the \emph{a priori} model can also be addressed in
a Bayesian context, by using a process of model selection
\cite{sivia93,loredo01} and defining an \emph{a posteriori}
distribution of parameters in the model \cite{jarrell96}. Further,
only single line profiles were analyzed here; while the formalism
has been expressed for overlapped line profiles, demonstrating
that Bayesian/MaxEnt method is flexible in its application,
additional analysis of overlapped line profiles is needed and this
will be followed in future studies.

The literature has seen considerable debate over the type of
distribution that best describes the distribution of sizes
\citeaffixed{blackman94,krill98,kiss99,langford00}{see}. In the
analysis presented here we have simply used a log-normal
distribution to demonstrate that the Bayesian/MaxEnt method can
reproduce the parameters. Moreover, the position we have taken in
developing the Bayesian/MaxEnt method is that we are not concerned
with the type of distribution; rather, we have produced a reliable
and consistent method that can determine the specimen profile
and/or the size distribution, given our understanding of the
experimental data, statistical noise and instrumental effects.

\end{document}